\documentclass[aps,prd,12pt,a4paper,groupedaddress,preprintnumbers,floatfix,nofootinbib]{revtex4}
\usepackage[english]{babel}
\usepackage{amsmath}
\usepackage{amsfonts}
\usepackage{amssymb}
\usepackage{subfig}
\usepackage{wrapfig}
\usepackage{graphicx}

\usepackage{color}
\usepackage{hyperref}
\usepackage{dcolumn}
\usepackage{bm}
\usepackage{float}
\newcommand{\be}{\begin{equation}}
\newcommand{\ee}{\end{equation}}
\newcommand{\bes}{\begin{subequations}}
\newcommand{\ees}{\end{subequations}}
\newcommand{\bea}{\begin{eqnarray}}
\newcommand{\eea}{\end{eqnarray}}
\newcommand{\bear}{\begin{equation}\begin{array}}
\newcommand{\eear}[1]{\end{array}\label{#1}\end{equation}}
\def\ba{$$\begin{array}}
\def\ea{\end{array}$$}
\def\bra{$\begin{array}}
 \def\era{\end{array}$}

 \newcommand{\rgg}{\mathcal{R}_{\gamma\gamma}}
 
\def\DS{\color{black}}

 \newcommand{\rzg}{\mathcal{R}_{Z\gamma}}
\newcommand{\g}{\,\mbox{GeV}}

\catcode`\@=11
\def\lsim{\mathrel{\mathpalette\@versim<}}
\def\gsim{\mathrel{\mathpalette\@versim>}}
\def\@versim#1#2{\vcenter{\offinterlineskip
\ialign{$\m@th#1\hfil##\hfil$\crcr#2\crcr\sim\crcr } }}
\catcode`\@=12

\catcode`@=12
\begin{document}

\title{IDMS: Inert Dark Matter Model with a complex singlet}

\author{
Cesar Bonilla,$^{1}$\footnote{Electronic address:cesar.bonilla@ific.uv.es}
Dorota Sokolowska,$^{2}$\footnote{Electronic address:Dorota.Sokolowska@fuw.edu.pl}
Neda Darvishi,$^{2}$\footnote{Electronic address:Neda.Darvishi@fuw.edu.pl}
J. Lorenzo Diaz-Cruz,$^{3}$\footnote{Electronic address:jldiaz@fcfm.buap.mx}
and Maria Krawczyk,$^{2}$\footnote{Electronic address:maria.krawczyk@fuw.edu.pl}}

\affiliation{
  $^1$ Instituto de F\'{i}ısica Corpuscular (CSIC-Universitat de Val\`{e}ncia), Apdo. 22085, E-46071 Valencia, Spain. \\
  $^2$ University of Warsaw, Faculty of Physics, Warsaw, Poland.\\
  $^3$ Facultad de Ciencias Fisico-Matematicas, Benemerita Universidad Autonoma de Puebla, Puebla, M\'exico.}  
  \date{\today}

\begin{abstract}
We study an extension of the Inert Doublet Model (IDM) that includes an extra complex singlet of the scalars fields, which we call the  IDMS.
In this model there are three Higgs particles, among them a SM-like Higgs particle, and the lightest neutral scalar, from the inert sector, remains a viable
dark matter candidate.
{We assume  a non-zero complex vacuum expectation value for the singlet, so that the visible sector can introduce extra sources of CP violation.}
We construct the scalar potential of IDMS, assuming an exact $Z_2$ symmetry,  with the new singlet being $Z_2$-even, as well as
a softly broken $U(1)$ symmetry, which allows a reduced number of { {free parameters in the potential}}.

 In this paper we explore the foundations of the model, in particular 
the masses and interactions of  scalar  particles for a few benchmark scenarios.  
Constraints from collider physics, in particular from the Higgs signal observed at LHC
with $M_h\approx 125$ GeV, as well as constraints from the dark matter experiments, such as relic density measurements and direct detection limits,
are included in the analysis. We observe significant differences with respect to the IDM in relic density values from additional annihilation channels, interference and resonance effects due to the extended Higgs sector.
\end{abstract}

\maketitle
\newpage
\baselineskip 24pt

\section{Introduction}
 After many years of expectations the LHC has found a Standard-Model-like (SM-like) Higgs particle
 with a mass of $M_h \approx 125$ GeV \cite{Aad:2012tfa,Chatrchyan:2012ufa}. Current analysis of LHC data has been
dedicated to the properties of this resonance, with the purpose of determining whether it 
belongs to the SM or to one of its extensions. In the later case some 
deviations from the SM predictions are expected.
The LHC has also provided important bounds on the scale
of new physics beyond the SM, either through the search for new (probably heavy) particles or by
looking for deviations from the SM predictions of properties of the SM particles.  Some of the motivations 
for new physics are related to cosmology, in particular the problem of dark matter (DM)
or the baryon asymmetry of the Universe (BAU).

 One of the simplest models for scalar dark matter is the Inert Doublet
Model (IDM), a version of a Two Higgs Doublet Model with an exact $Z_2$ symmetry \cite{Deshpande:1977rw}. 
Here the SM scalar (Higgs) sector is extended by an inert scalar doublet.
This model can account for a SM-like Higgs particle, and at the same time for the correct 
relic density of dark matter, while fulfilling  direct and indirect DM detection limits, while simultaneously agreeing 
with the LHC results [see e.g. \cite{Ma:2006km,Barbieri:2006dq,LopezHonorez:2006gr,Goudelis:2013uca, Krawczyk:2013jta,Arhrib:2013ela}].

Furthermore, the IDM can provide a strong first-order phase transition \cite{Gil:2012ya},
 which  is a { desired}  condition needed to generate a
baryon asymmetry of the Universe. However,
the IDM contains no additional source of CP violation with respect to the SM, and the only CPV phase comes 
from the CKM matrix, as in the SM, which is known to be too small to lead to the right amount of BAU.

In this paper we shall extend the IDM by including a complex scalar singlet $\chi$, which accompanies the 
SM-like Higgs doublet and  inert doublet, denoted here by $\Phi_1$ and $\Phi_2$, respectively. We shall call this 
model the IDMS (the IDM plus singlet). 
{ A complex
non-zero  vacuum expectation value for the singlet field is assumed}.
Our main aim is to study general properties of the model, and to check its agreement with all existing Higgs- and DM data. 
We expect that the extended Higgs sector will influence DM annihilation, as well as its detection prospects, due to new CP-violating annihilation channels, interference between new diagrams and resonance effects.
Although a detailed investigation of the CP violating effects  is beyond the scope of this paper, we hope
to lay the foundations for a model that is consistent both from theoretical and phenomenological constraints, where
such aspects could be studied consistently in the future. 

The content of this paper is as follows. Section \ref{sec-gen} contains the presentation
of the general model, in particular its  scalar potential. In section \ref{sec-our} we present in detail a constrained
version of our model, including positivity conditions,  the mass eigenstates in the neutral and charged 
sectors and study the parameter space of the model. Section \ref{sec-lhc} contains an analysis of Higgs couplings and a comparison with
LHC data. In section \ref{sec-dm} we present our study of relic density for a dark matter candidate
of the model, which is assumed to be the lightest neutral $Z_2$-odd scalar state. Conclusions are presented in section \ref{sec-con}, 
where we also discuss possible implications for neutrino physics. Detailed formulas, benchmark points and values related
to the LHC and dark matter analysis are presented in the appendices.

\section{The IDMS: The IDM plus a complex singlet \label{sec-gen}}

We shall consider a $Z_2$-symmetric model that contains a SM-like Higgs doublet $\Phi_1$, which is involved in a generation of
the masses of gauge bosons and fermions, as in the SM. There is also an inert scalar doublet $\Phi_2$,
which is odd under a $Z_2$ symmetry. This $\Phi_2$ doublet  has VEV$=0$ and can provide a stable dark matter candidate. Then, we have the neutral
complex singlet $\chi$ with hypercharge $Y=0$  and a non-zero complex VEV.

The singlet $\chi$ can play several roles in  models with two doublets and a singlet, leading to different scenarios. CP violation can be explicit, provided by the singlet interaction terms, or spontaneous, if $\langle \chi \rangle \in \mathbb{C}$.

The singlet $\chi$ could be even or odd under a $Z_2$ symmetry, and it could
mix with the SM-like Higgs doublet and/or with the inert doublet. Furthermore, one could even
use the complex singlet to induce all sources of CP violation, including the SM one contained
in the CKM mixing matrix, as it was done in Ref. \cite{Branco:2003rt}.

Here we shall take $\chi$  to be even under a $Z_2$ transformation defined as:
\begin{equation}
Z_2\;:\; \Phi_1 \to \Phi_1, \;  \Phi_2 \to - \Phi_2, \; \textrm{SM fields} \to  \textrm{SM fields}, \; \chi \to \chi, \label{z2}
\end{equation}
and allow its mixing only with the neutral  components of $\Phi_1$; furthermore, we shall consider the case when the 
CP symmetry can be violated  by a non-zero complex $\langle \chi \rangle$.

The full Lagrangian of the model looks as follows:
\begin{equation}
{ \cal L}={ \cal L}^{SM}_{ gf } +{ \cal L}_{scalar} + {\cal L}_Y(\psi_f,\Phi_{1}) \,, \quad { \cal L}_{scalar}=T-V\, ,
\label{lagrbas}
\end{equation}
where ${\cal L}^{SM}_{gf}$ describes boson-fermion interaction as in the SM, ${ \cal L}_{scalar}$ describes the scalar 
sector of the model, and ${\cal L}_Y(\psi_f,\Phi_{1})$ -- the Yukawa interaction. The kinetic term in ${ \cal L}_{scalar}$ 
has the standard form:
\begin{eqnarray}
T = \left( D_{\mu} \Phi_{1} \right)^{\dagger} \left( D^{\mu} \Phi_{1} \right) + \left( D_{\mu} \Phi_{2} 
\right)^{\dagger} \left( D^{\mu} \Phi_{2} \right) + \partial \chi \partial \chi^*, \label{kinet}
\end{eqnarray}
with $D^\mu$ being a covariant derivative for an $SU(2)$ doublet. 

We take the Yukawa interaction in the form of the  Model I in the 2HDM, where  only $\Phi_1$ couples to fermions.

Within our model the scalar singlet $\chi$  does not couple with the SM fermions and therefore the singlet-fermion 
interaction are present only through mixing of singlet with the first doublet  $\Phi_1$.

 In our model only $Z_2$-even fields $\Phi_1$ and $\chi$ acquire vacuum expectation values, which we denote by $v$ and $w e^{i\xi}$, 
 respectively, where  $v,w, \xi \in \mathbb{R}$. We shall use the following field decomposition around the 
 vacuum state $(v,0,w e^{i\xi})$:

\begin{eqnarray}
& \Phi_{1} = \left( \begin{array}{c} \phi^+_1 \\ \frac{1}{\sqrt{2}} \left( v + \phi_1 + i \phi_6 \right)\\ 
\end{array} \right), \qquad \Phi_{2} = \left( \begin{array}{c}\phi^+_2 \\ \frac{1}{\sqrt{2}} \left(  \phi_4 + 
i \phi_5 \right)\\ \end{array} \right), \label{dec_doublets}&\\[2mm]
& \chi = \frac{1}{\sqrt{2}} (w e^{i \xi} + \phi_2 + i \phi_3). \label{dec_singlet}&
\end{eqnarray}

Thus, the $Z_2$ symmetry (\ref{z2}) is  not violated spontaneously. Also, $U(1)_{EM}$ 
is not broken,  and  there is no mixing between the neutral and charged components. Masses of gauge bosons and 
fermions are given by the VEV of the first doublet as in the SM,  e.g $M_W^2 = g^2 v^2/4$ for the $W$ boson.


The full scalar potential of the model can be written as
\begin{equation}
V=V_{IDM}+V_S+V_{DS}, \label{potgen}
\end{equation}
where we have separated the pure doublet and the pure singlet parts (respectively $V_{IDM}$ and $V_{S}$) 
and their interaction term ($V_{DS}$).
The IDM part of the potential, $V_{IDM}$, is given by:
\bea
  &
    \begin{array}{c}
V_{IDM} = - \frac{1}{2}\left[{m_{11}^2} \Phi_1^\dagger\Phi_1 + {m_{22}^2} \Phi_2^\dagger\Phi_2 \right] 
+ \frac{1}{2}\left[\lambda_1 \left(\Phi_1^\dagger\Phi_1\right)^2 
+ \lambda_2 \left(\Phi_2^\dagger\Phi_2 \right)^2\right]\\[6mm]
+  \lambda_3 \left(\Phi_1^\dagger\Phi_1 \right) \left(\Phi_2^\dagger\Phi_2\right) + \lambda_4 \left(\Phi_1^\dagger\Phi_2\right) 
\left(\Phi_2^\dagger\Phi_1\right) +\frac{\lambda_5}{2}\left[\left(\Phi_1^\dagger\Phi_2\right)^2\!+\!\left(\Phi_2^\dagger\Phi_1\right)^2\right].
    \end{array}&
\label{potIDM}
\eea

The general singlet part of the potential  is equal to:
\bea
  &
    \begin{array}{c}
V_{S} = -\frac{m_3^2}{2} \chi^* \chi -\frac{m_4^2}{2} (\chi^{*2} + \chi^2) + \lambda_{s1} (\chi^*\chi)^2 + \lambda_{s2} 
(\chi^*\chi)(\chi^{*2} + \chi^2) + \lambda_{s3} (\chi^4 + \chi^{*4})\\[2mm]
+ \kappa_1 (\chi + \chi^*) + \kappa_2 (\chi^3 + \chi^{*3}) + \kappa_3( \chi(\chi^*\chi) + \chi^*(\chi^*\chi)).
    \end{array}&
\label{potS}
\eea
The doublet-singlet interaction terms are:
\bea
  &
    \begin{array}{c}
V_{DS} = \Lambda_1(\Phi_1^\dagger\Phi_1)(\chi^* \chi) + \Lambda_2 (\Phi_2^\dagger\Phi_2)(\chi^* \chi) +  
\Lambda_3 (\Phi_1^\dagger\Phi_1)(\chi^{*2}+ \chi^2) + \Lambda_4 (\Phi_2^\dagger\Phi_2)(\chi^{*2} +\chi^2)\\[2mm]
+ \kappa_4 (\Phi_1^\dagger\Phi_1) ( \chi +\chi^*) + \kappa_5 (\Phi_2^\dagger\Phi_2) ( \chi +\chi^*).  \end{array}&
\label{potMix}
\eea

We assume that all parameters of $V$ (\ref{potgen}) are real, and it is not difficult to see that the potential is explicitly invariant under a CP transformation $\Phi_{1,2} \to \Phi_{1,2}^\dagger, \chi \to \chi^\star$.

As $V$ is $Z_2$-symmetric and the chosen vacuum state (\ref{dec_doublets},\ref{dec_singlet}) will not 
spontaneously break this symmetry, the problem of cosmological domain walls will not arise in this model. 
In total, there are four quadratic parameters, twelve dimensionless quartic parameters and five dimensionful 
parameters $\kappa_{1,2,3,4,5}$. The linear term $\kappa_1$ can be removed by a translation of the singlet field, and 
we will omit it below.

 One could reduce this general model by invoking
additional symmetries besides the imposed $Z_2$ one (see e.g. \cite{Kadastik:2009dj, Kadastik:2009cu, Kadastik:2009ca, Belanger:2012vp, Belanger:2014bga, Banik:2014cfa, Banik:2014eda, Curtin:2014jma} for various symmetry assignments). In particular, to simplify the model
one can apply a global $U(1)$ symmetry, as we discuss below. Here only the scalars from the inert doublet may be considered 
as dark matter candidate - in contrast to the fermion singlet being charged, see \cite{Mambrini:2015sia}. Similarly,
had we chosen to  assign a $Z_2$-odd quantum number also to $\chi$ (or if singlet was odd under an additional 
$Z_2'$ symmetry), it would have also resulted in a variant 
of the model with a simplified potential, where all terms with an odd number of field $\chi$ would be absent. 
Obviously, in those cases having a $Z_2$ (or $Z_2'$) symmetric vacuum state would require $\langle \chi \rangle =0$, 
and thus there would be no additional CP violation in the model.

\section{The constrained IDMS: cIDMS \label{sec-our}}
We will reduce the most general IDMS  potential (\ref{potgen}-\ref{potMix})
by imposing  a global $U(1)$ symmetry: 
\begin{equation}
U(1): \;\; \Phi_1 \to \Phi_1,\, \Phi_2 \to \Phi_2, \, \chi \to e^{i\alpha} \chi \label{u1def}.
\end{equation}

However, a non-zero VEV $\langle \chi \rangle$ would lead to a spontaneous breaking of this continuous symmetry and
appearance of massless Nambu-Goldstone scalar particles, which are not phenomenologically viable. 
Keeping some $U(1)$-soft-breaking terms in the potential would solve this problem and at the same time 
would still lead to a reduction of the  number of parameters in $V$.

{ The parameters of the IDMS potential can be divided into the following groups:
\begin{enumerate}
\item $U(1)$-symmetric terms: $m_{11}^2, m_{22}^2, m_{3}^2, \lambda_{1,2,3,4,5}, \lambda_{s1}, \Lambda_{1,2}$,
\item $U(1)$-soft-breaking terms\footnote{Recall that $\kappa_1$ can be removed from (\ref{potgen}) 
by translation of $\chi$.}: $m_{4}^2, \kappa_{2,3}, \kappa_{4,5}$,
\item $U(1)$-hard-breaking terms $\lambda_{s2}, \lambda_{s3}, \Lambda_{3,4}$.
\end{enumerate}

In what follows we shall consider a potential with soft-breaking of the $U(1)$ symmetry by the singlet 
cubic terms $\kappa_{2,3}$ and quadratic term $m_4^2$ only,  neglecting the remaining ones ($\kappa_{4,5}$).  
We recall that $\Phi_1$ is the SM-like Higgs doublet responsible 
for the EW symmetry breaking and for providing masses of gauge bosons and fermions. In addition, 
we want to use it as a
portal for DM interactions with the visible sector, as in the IDM.   We shall assume 
therefore that there is no direct coupling 
of $\Phi_2$ to $\chi$, thus setting the $U(1)$-invariant term $\Lambda_2 = 0$. 
The field  $\chi$ shall then interact 
with the DM particles only through mixing with the neutral component of $\Phi_1$.

We are therefore left with the following $U(1)$-symmetric terms
($m_{11}^2, m_{22}^2, m_{3}^2, \lambda_{1-5}, \lambda_{s1}, \Lambda_{1}$) and $U(1)$-soft-breaking 
terms ($m_{4}^2, \kappa_{2,3}$). }

We shall call our model,  the model
with this choice of parameters, cIDMS. {The cIDMS potential  is then given by:}
\bea
  &
    \begin{array}{c}
V = -\frac{1}{2}\left[{m_{11}^2} \Phi_1^\dagger\Phi_1 + {m_{22}^2} \Phi_2^\dagger\Phi_2 \right] 
+ \frac{1}{2}\left[\lambda_1 \left(\Phi_1^\dagger\Phi_1\right)^2 
+ \lambda_2 \left(\Phi_2^\dagger\Phi_2 \right)^2\right]\\[6mm]
+  \lambda_3 \left(\Phi_1^\dagger\Phi_1 \right) \left(\Phi_2^\dagger\Phi_2\right) + \lambda_4 
\left(\Phi_1^\dagger\Phi_2\right) \left(\Phi_2^\dagger\Phi_1\right) +\frac{\lambda_5}{2}\left[\left(\Phi_1^\dagger\Phi_2\right)^2\!
+\!\left(\Phi_2^\dagger\Phi_1\right)^2\right] \\[3mm]
-\frac{m_3^2}{2} \chi^* \chi + \lambda_{s1} (\chi^*\chi)^2 + \Lambda_1(\Phi_1^\dagger\Phi_1)(\chi^* \chi)\\[2mm]
 -\frac{m_4^2}{2} (\chi^{*2} + \chi^2) + \kappa_2 (\chi^3 + \chi^{*3}) + \kappa_3 [ \chi(\chi^*\chi) + \chi^*(\chi^*\chi)].
    \end{array}&
\label{potIDM1S}
\eea

\subsection{Comments about parameter choice}

Once the potential  (\ref{potgen}) is restricted only to $U(1)$-symmetric or $U(1)$-soft-breaking terms, no more terms will be generated when we move beyond tree-level. For our choice of parameters, the cIDMS, we assume that some of $U(1)$-symmetric or $U(1)$-soft-breaking terms are set manually to zero. One may ask these terms will remain zero, or if they will be generated at loop level. Indeed, it turns out that some terms we neglected, namely 
$\kappa_4$ and $\Lambda_2$ are generated already at the 1-loop level, with their $\beta$ functions being proportional to  $ \frac{1}{16 \pi^2}$ and product of $ 
\Lambda_1$ and  $\lambda_3$ and, respectively, a combination of $ \lambda_4, \kappa_3$
\cite{marcosampaio}.

This shows that our parameter choice is not protected against loop corrections, which was expected, as those terms are allowed by the symmetry we chose to consider. However, it is important to notice that loop contributions for both $\kappa_4$ and $\Lambda_1$ depend on the parameter $\Lambda_1$, i.e. the mixing parameter between $\Phi_1$ and $\chi$. In our analysis we chose scenarios where this parameter is small, leading to the Higgs particle being SM-like, which is a favoured interpretation of current LHC data.\footnote{The linear term,  with $\beta$ function  $ \propto 1/{16 \pi^2}(m_3^2 \kappa_3 + m_4^2(3\kappa_2+\kappa_3))$,  
even if removed by translation of fields at tree-level, appears when we include loop corrections. The resulting tadpole diagram can be interpreted as the shift in vacuum energy. If $\kappa_1$ is kept non-zero at tree-level, one can remove it consistently at every loop level \cite{linear}. In any case, this term is not relevant for the presented work.}  

One can notice also that if $\kappa_3$ is equal to zero, then both $\kappa_4$ and $\Lambda_2$ remain zero also at loop level. This, and other parameter choices, are left for the future work \cite{inprogress}.

\subsection{Positivity conditions}

In order to have a stable minimum, the parameters of the potential need to
satisfy positivity conditions. Namely, the potential should be bounded from below, i.e. should not go to negative
infinity for large field values. As this behaviour is dominated by the quartic terms, the cubic terms will 
not play a role here. Thus the following conditions will apply to 
a variety of  models that will differ only by their cubic interactions. 

We use the method of \cite{Kannike:2012pe}, which uses the concept of co-positivity for a matrix build of 
coefficients in the field directions. For the  cIDMS, the 
 positivity conditions read:

\begin{equation}
\begin{array}{l}
 \lambda_1, \lambda_2, \lambda_{s1} \geq 0, \; {\bar{\lambda}_{12}} = \lambda_3 + \theta[-\lambda_4+|\lambda_5|] 
 (\lambda_4-|\lambda_5|) + \sqrt{\lambda_1 \lambda_2}  > 0, \\[3mm]
 {\bar{\lambda}_{1S}} = \Lambda_1  + \sqrt{2 \lambda_1 \lambda_{s1}} > 0,\\[3mm]
 \frac{1}{2} \sqrt{\lambda_1 \lambda_2\lambda_{s1}} + 
[\lambda_3 + \theta[-\lambda_4+|\lambda_5|] (\lambda_4-|\lambda_5|)] \sqrt{\lambda_{s1}}+
\Lambda_1 \sqrt{ \frac{\lambda_2}{2}} 
+ \sqrt{ {\bar{\lambda}_{12}} {\bar{\lambda}_{1S}} {\bar{\lambda}_{2S}} } > 0,
\end{array} \label{pos}
\end{equation}
where ${\bar{\lambda}_{2S}} =  \sqrt{2 \lambda_2 \lambda_{s1}} >0$.

\subsection{Extremum conditions} 
It is useful to re-express dimensionful parameters $\kappa_{2,3}$ in terms of the dimensionless parameters $\rho_{2,3}$
(we consider them being of order $\mathcal{O}(1)$) as:
\begin{equation}
\kappa_{2,3} = w \rho_{2,3},
\end{equation}
with $w$ being an absolute value of the singlet VEV.

The minimization conditions lead to the following constraints for three  quadratic 
parameters from $V$ (\ref{potIDM1S}):
\begin{eqnarray}
&& m_{11}^2 =  w^2 \Lambda_1 + v^2 \lambda_1, \\
&& m_{3}^2 = v^2 \Lambda_1 + 2w^2\lambda_{s1}+\frac{w^2}{\sqrt{2}\cos\xi}(-3 \rho_2 + 3\rho_3 + 2\rho_3 \cos2\xi),\\
&& m_4^2 =  \frac{w^2}{2\sqrt{2}\cos\xi} (3 \rho_2 + \rho_3 + 6\rho_2 \cos2\xi).
\end{eqnarray}

The $m_{22}^2$ parameter is not determined by the extremum conditions,  just like in the IDM.

The squared-mass matrix $M_{ij}^2$, for $i,j=1,...6$,  is given by:
\begin{equation}
M_{ij}^{2} = \frac{\partial^{2} V}{\partial \phi_{i} \phi_{j}}\biggr\vert_{\Phi_{i} =\left\langle \Phi_{i} 
\right\rangle,\chi = \left\langle \chi \right\rangle}, \label{massmat_def}
\end{equation}
with $\phi_i$ being the respective fields from the decomposition (\ref{dec_doublets},\ref{dec_singlet}).
This definition along with the normalization defined in (\ref{dec_doublets},\ref{dec_singlet}) gives the proper 
mass terms of $M_\varphi^2 \varphi^+ \varphi^-$ for the charged scalar fields, and 
$\frac{M_\varphi^2}{2} \varphi^2$ for the neutral scalar fields. 

\subsection{Comments on vacuum stability}
The tree-level positivity conditions (\ref{pos}), which ensure the existence of a global minimum, correspond to $\lambda>0$ 
in the Standard Model. It is well known, that radiative corrections 
coming from the top quark contribution can lead to negative values of the Higgs self-coupling, resulting in the 
instability of the SM vacuum for larger energy scales. Full analysis of the stability of the cIDMS potential beyond 
 tree-level   is beyond the scope of this paper. However, it has been shown in  a simple approach based on the tree-level
 condition for vacuum stability that for the IDM  the contributions from additional scalar states will in general 
lead to the relaxation of the stability bound at high energies and allow the IDM to be valid up to the Planck scale Ref.~\cite{Goudelis:2013uca}. Since cIDMS 
contains two more scalar states, in principle this condition should hold here as well. 
However, one should keep in mind that a treatment within the effective potential approach is needed in order to study this aspect in detail.
\subsection{Mass eigenstates }
\subsubsection{The neutral sector \label{mass-neutral-intro}}
The form of the neutral part of the squared-mass matrix (\ref{massmat_def}) for $\phi_i, (i=1,...,6)$ allows us to 
identify the physical states and their properties: 
\begin{equation}
M^2 = \left(
\begin{array}{cc}
M_{mix(3\times3)}^2 & 0_{(3\times3)} \\
0_{(3\times3)} & \begin{array}{ccc}
M_H^2 & 0 & 0 \\
0 & M_A^2 & 0 \\
0 & 0 & 0 \\
\end{array} \\
\end{array}
\right)\label{mass_neutral}
\end{equation}
As there is no mixing between four $Z_2$-even fields $\phi_{1,2,3,6}$, and two $Z_2$-odd fields $\phi_{4,5}$, we can divide the particle content of the model into two separate sectors: 
the $Z_2$-even sector, called \textit{the Higgs sector}, and the $Z_2$-odd sector, called \textit{the inert sector}. Below we list the particle content of the neutral sector:
\begin{enumerate}
\item The Goldstone field, $G_Z = \phi_6$, is a purely imaginary part of the first doublet $\Phi_1$.
\item There is a mixing between the singlet $\chi$ and the 
real neutral fields of $\Phi_1$ (namely $\phi_1, \phi_2$ and $\phi_3$)  resulting in  three neutral scalars $h_1, h_2, h_3$.
 Due to the non-zero phase of the singlet 
VEV ($w e^{i\xi}$) the fields  $h_1, h_2, h_3$  are  composed  of  states of different CP properties. Therefore  among 
the possible vertices there are vertices like $Z Z h_i$ and  all $h_i$ particles couple to fermions. 
Masses of the these Higgs particles depend only on the following parameters of the potential: $\lambda_1, \Lambda_1, \rho_{2,3}, \lambda_{s1}$.
\item In the inert sector the dark matter candidate from the IDM is  stable and it is the 
lighter of the two neutral components of $\Phi_2$ ($ \phi_4$ or $ \phi_5$), which we identify 
as the scalar particles $H$ and $A$. Masses of those particles are just like in the IDM: 
\begin{eqnarray} 
&& M_H^2 = \frac{1}{2} ( -m_{22}^2 + v^2 \lambda_{345}) , \quad \; H = \phi_4,\\
&& M_A^2 = \frac{1}{2} ( -m_{22}^2 + v^2 \lambda_{345}^-), \quad \; A = \phi_5,
\end{eqnarray}
where $\lambda_{345} = \lambda_3 + \lambda_4 + \lambda_5$, $\lambda_{345}^- = \lambda_3 + \lambda_4 - \lambda_5$. 
Notice, that the IDM relation for masses still holds:
\begin{equation}
\lambda_5 = \frac{M_H^2-M_A^2}{v^2}. \label{lam5IDM}
\end{equation}
If $\lambda_5 < 0$ then $H$, as a neutral lighter state, is  our  dark matter candidate. 
Since $Z_2$ symmetry is exact in our model,  the $Z_2$-odd particles have limited gauge 
and scalar interactions (they interact in pairs only) and they do not couple to fermions.
Masses of inert particles (also charged scalars) depend only on $\lambda_{3,4,5}$ and  $m_{22}^2$.
These parameters do not influence masses of the Higgs particles from the $Z_2$-even sector. 
In that sense, the masses of particles from the Higgs and inert sectors can be studied separately. 
On this level, the only connection between parameters from these two sectors is through the positivity constraints.
As in the IDM, $\lambda_2$ does not influence the mass sector and it appears only as a quartic 
coupling between the $Z_2$-odd particles.
\end{enumerate}

\subsubsection{The charged sector}
The $Z_2$-odd charged scalar $H^\pm$ comes 
solely from the second doublet, as in the IDM; its mass is given by 
\begin{equation}
M_{H^\pm}^2 = \frac{1}{2} (- m_{22}^2 + v^2 \lambda_3).
\end{equation}

Notice, that the mass relations for the $Z_2$-odd sector from the IDM  hold, namely 
\begin{equation}
M_H^2 = M_{H^\pm}^2 + \frac{v^2 (\lambda_4 + \lambda_5)}{2}, \quad M_A^2 = M_{H^\pm}^2 + 
\frac{v^2 (\lambda_4 - \lambda_5)}{2}. \label{relIDM}
\end{equation}
The neutral particle $H$ is a DM candidate, therefore $\lambda_4+\lambda_5 <0$, resulting in $M_H < M_{H^\pm}$.

If we allow an additional mixing between $\Phi_2$ and $\chi$ through a non-zero  $\Lambda_{2,4}$ and
 $\rho_5$ then the squared-mass formulas are modified as $M_{H,A,H^\pm}^2 \to M_{H,A,H^\pm}^2 + \Delta$, with 
$
\Delta= \frac{1}{2} w^2 (\Lambda_2 + 2  \Lambda_4 \cos2\xi + 2\sqrt{2} \rho_5 \cos\xi ).\nonumber
$
Still, the IDM relations (\ref{lam5IDM}) and (\ref{relIDM}) hold.

\subsection{Physical states in the Higgs sector \label{ssec-phys}}

The mass matrix that describes the singlet-doublet mixing, in the basis of neutral fields $(\phi_1, \phi_2, \phi_3)$, 
is given by: 
\begin{equation}
M^2_{mix} = \left(
\begin{array}{ccc}
\mu_{11} & \mu_{12} & \mu_{13}\\
\mu_{12} & \mu_{22} & \mu_{23}\\
\mu_{13} & \mu_{23} & \mu_{33}
\end{array} \right),\label{massneut}
\end{equation}
 where matrix elements $\mu_{ij}$ are
 \begin{eqnarray}
 &&\mu_{11} =  \lambda_1 v^2, \label{eq25}\\
 &&\mu_{12}= w v \Lambda_1  \cos \xi, \\
 &&\mu_{13}= w v \Lambda_1  \sin \xi, \\
 &&\mu_{22} = \frac{w^2}{2 \cos \xi}  \left(3\sqrt{2} \rho_2 + \sqrt{2}\rho_3(1+ 2\cos2\xi) + 
 \lambda_{s1} (3\cos\xi+ \cos3\xi)  \right), \label{eq28}\\
 &&\mu_{23} = w^2 \left(\sqrt{2}(-3\rho_2+\rho_3)+2 \lambda_{s1} \cos\xi \right) \sin \xi,\\
 &&\mu_{33} =2 w^2 \sin^2 \xi \lambda_{s1}. \label{matrixel}
 \end{eqnarray}
Only when  $\Lambda_1 \not=0$ and $w, \, \sin \xi  \not =0$, there is a mixing between states of 
different CP properties   $\phi_1$ or $\phi_2$ and $\phi_3$ (entries $\mu_{13}$ and $\mu_{23}$ respectively).

Diagonalization of $M^2_{mix}$ (\ref{massneut}) gives the mass eigenstates, which can be also obtained by 
the rotation of the field basis:
\begin{eqnarray}
 & \left( \begin{array}{c} h_1\\ h_2\\ h_3\\ \end{array} \right) = R \left( \begin{array}{c} 
 \phi_1\\ \phi_2\\ \phi_3\\ \end{array} \right), \quad \widetilde{M}^2 = R M_{mix}^2 R^T = diag(M_{h_1}^2,M_{h_2}^2,M_{h_3}^2). & \label{neutr_diag}
\end{eqnarray}
The rotation matrix $R = R_1 R_2 R_3$ in principle depends on  three mixing angles ($\alpha_1, \alpha_2, \alpha_3$).  
The individual rotation matrices are given by (here and below $c_i = \cos \alpha_i, s_i = \sin \alpha_i$):
\begin{equation}
R_1 = \left(
\begin{array}{ccc}
c_1 & s_1 & 0\\
-s_1 & c_1 & 0\\
0 & 0 & 1
\end{array} \right), \quad R_2 = \left(
\begin{array}{ccc}
c_2 & 0 & s_2\\
0 & 1 & 0\\
-s_2 & 0 & c_2
\end{array} \right), \label{gen_rot}
\end{equation}
and
\begin{equation}
R_3 = \left(
\begin{array}{ccc}
1 & 0 & 0\\
0 & c_3 & s_3\\
0 & -s_3 & c_3
\end{array} \right).\label{rot23}
\end{equation}
 All $\alpha_i$ vary over an interval of length $\pi$.
The full rotation matrix depends on the mixing angles in the following way:
\begin{equation}
R = R_1 R_2 R_3 = \left(
\begin{array}{ccc}
c_1 c_2 & c_3 s_1 - c_1 s_2 s_3 & c_1 c_3 s_2 + s_1 s_3\\
-c_2 s_1 & c_1 c_3 + s_1 s_2 s_3 & -c_3 s_1 s_2 + c_1 s_3\\
-s_2 &  -c_2 s_3 & c_2 c_3
\end{array} \right).\label{rotfull}
\end{equation}
The inverse of $R$ can be used to obtain the reverse relation between $h_i$ and $\phi_i$:
\begin{equation}
R^{-1}= \left(
\begin{array}{ccc}
c_1 c_2 &  -c_2 s_1 & -s_2\\
c_3 s_1 - c_1 s_2 s_3 & c_1 c_3 + s_1 s_2 s_3&-c_2 s_3\\
c_1 c_3 s_2 + s_1 s_3   & -c_3 s_1 s_2 + c_1 s_3 & c_2 c_3
\end{array} \right)\label{rotinvfull}.
\end{equation}
The two important relations   can be read from these rotation matrices, namely:
\begin{equation}
h_1 = c_1 c_2 \phi_1 + (c_3 s_1 - c_1 s_2 s_3) \phi_2 + (c_1 c_3 s_2 + s_1 s_3) \phi_3
\end{equation}
and
\begin{equation}
\phi_1 = c_1 c_2 h_1 - c_2 s_1 h_2 - s_2 h_3 \label{phi1DM}.
\end{equation}
The above equations describe the composition of the SM-like Higgs boson $h_1$, in terms of real components  $\phi_1$ and  $\phi_2$, 
which provide a CP-even part, as well as the $\phi_3$ component -- CP-odd one.
Equivalently, one can look at it as the modification of 
the real component of the SM-like Higgs doublet $\Phi_1$ from the cIDMS with respect to the SM and the IDM. 

Especially important is the first element both in $R$ and $R^{-1}$ equal to:
\begin{equation}
R_{11} = R^{-1}_{11} = c_1 c_2. \label{r11}
\end{equation}
This matrix element gives the relative modification of the interaction of the Higgs boson ($h_1$) with respect to the IDM, and will be important both in the LHC analysis (section \ref{sec-lhc}), and in the DM studies (sec. \ref{sec-dm}).

\subsection{Parameter space in the Higgs sector \label{ssec-cor}}

In what follows we shall numerically analyze the allowed regions of the parameter space of our model. In scans the positivity (\ref{pos}) and 
perturbativity conditions, where all quartic parameters in the potential are taken to be below 1, are fulfilled.

As LHC data is favouring a SM-like interpretation of the observed 125 GeV Higgs signal, we shall
require that the lightest neutral Higgs state comes predominantly from the doublet $\Phi_1$. If 
there was no $\Phi_1 - \chi$ mixing, then the SM-like Higgs boson's mass would have been given by $M_{h_1}^2 = v^2 \lambda_1 \Rightarrow \lambda_1 \approx 0.23$ (for $v =246$ GeV). We are going to consider the variation of $\lambda_1$ in range: 
\begin{eqnarray}
 0.2 < \lambda_1 < 0.3,\label{ll}
\end{eqnarray}
and demand that the mass of the lightest Higgs particle $h_1$ lies in range\footnote{The considered mass range [124.69, 125,37] GeV is in the 2$\sigma$ range in agreement with the newest LHC data \cite{Khachatryan:2014ira, Aad:2014eva} for the Higgs mass.}:
\be 
M_{h_1} \in  [124.69, 125.37] {\rm \, GeV}. \label{mh}
\ee
 The additional two Higgs scalars are heavier, we take
\be
M_{h_3} > M_{h_2} > 150\, \g. \label{mm}
\ee

Remaining parameters of the Higgs sector change in the following ranges:
\begin{eqnarray}\label{newlimits}
&&-1 < \Lambda_1 <1, \quad 0 < \lambda_{s1} < 1, \quad -1 <\rho_{2,3} < 1, \quad 0 <\xi < 2 \pi.
\end{eqnarray}

The parameters describing the inert sector, i.e. $\lambda_{2-5}, m_{22}^2$, do not directly influence values 
of masses of Higgs particles (\ref{massneut}-\ref{matrixel}). One must remember however, 
that allowed values of $\lambda_{2-5}$ are related to the ranges of Higgs parameters through 
the positivity constraints (\ref{pos}). In the scans, inert parameters change in the range allowed by 
the perturbativity constraints, with $H$ being the DM candidate (see sec. \ref{ssec-cordm}):
\begin{equation}
0<\lambda_2<1, \quad -1<\lambda_{3,4} < 1, \quad -1 <\lambda_5<0.
\end{equation}

We performed the scanning for $w \sim v=246 \g$, in particular for $w = 300, 500, 1000 \g$.  
{However, after noting that the results} do not depend strongly on the exact value of this 
parameter, {we opted here to present results with plots only} for $w=300 \g$.

In figures 1,2 and 3 correlations between parameters of the potential  related to the Higgs sector are shown.

\begin{figure}[H]
\vspace{-10pt}
  \centering
  \subfloat[$(\lambda_{s1}, \Lambda_1)$]{\label{ls1L1}\includegraphics[width=0.45\textwidth]{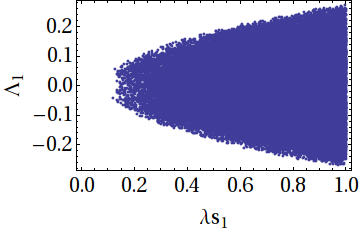}} \quad
  \subfloat[$(\Lambda_{1}, \xi)$]{\label{L1xi}\includegraphics[width=0.45\textwidth]{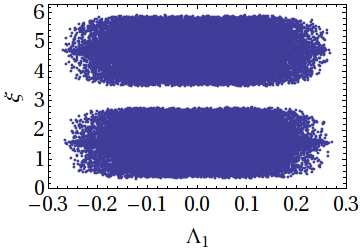}} \\
\vspace{-5pt}
  \caption{Correlations between parameters in the Higgs sector. Results of scanning for $w = 300$ GeV, 
  with ranges of parameters defined by eqs. (\ref{ll}-\ref{newlimits}).  \label{scan1}}
\end{figure}

\begin{figure}[H]
\vspace{-10pt}
  \centering
  \subfloat[$( \lambda_{s1},\rho_2)$]{\label{r2ls1}\includegraphics[width=0.45\textwidth]{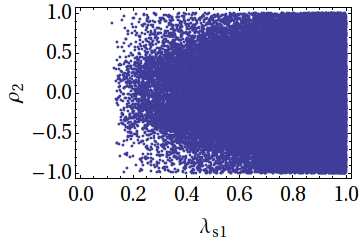}} \quad
  \subfloat[$(\rho_2, \rho_3)$]{\label{r2r3}\includegraphics[width=0.45\textwidth]{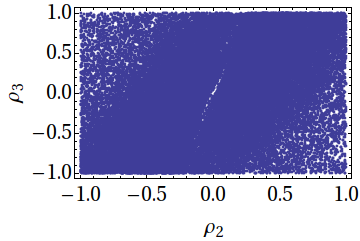}}\\
\vspace{-5pt}
  \caption{Correlations between parameters in the Higgs sector. Results of scanning for $w = 300$ GeV, 
  with ranges of parameters defined by eqs. (\ref{ll}-\ref{newlimits}). \label{scan1b}}
\end{figure}

$\bullet$ Fig. \ref{ls1L1} and \ref{r2ls1} show the allowed regions
in the planes $(\lambda_{s1},\Lambda_1)$ and $(\lambda_{s1},\rho_2)$. Notice the limited range of $\Lambda_1$ 
and the lower limit for $\lambda_{s1} \sim 0.1$. Both limits are arising from the mass ranges used 
in the scan. The positivity condition leads to the lower bound  on the negative $\Lambda_1$ only, however 
it is much weaker than the constraints coming  from the assumed limits on masses.

$\bullet$ Results of scanning presented in Figs. \ref{ls1L1}, \ref {L1xi} and \ref{r2ls1}  show that the range 
of $\Lambda_1$ is  limited with respect to the initial assumptions (\ref{newlimits}), and that good solutions 
require $|\Lambda_1|\lesssim 0.25$. Recall that this parameter describes mixing between $\Phi_1$ and $\chi$, 
effectively giving the non-SM contribution to the SM-like Higgs doublet.

$\bullet$ There is no correlation between $\rho_2$ and $\rho_3$, what is seen on Fig. \ref{r2r3}, where
  points in the $(\rho_2,\rho_3)$ plane are almost uniformly distributed.
 
\begin{figure}[H]
\vspace{-10pt}
  \centering
  \subfloat[$(\rho_{2}, \xi)$]{\label{r2xi}\includegraphics[width=0.45\textwidth]{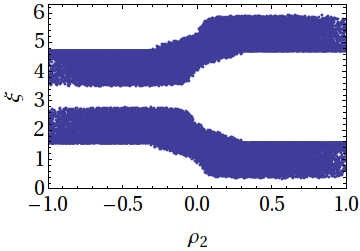}} \quad
  \subfloat[$(\rho_{3}, \xi)$]{\label{r3xi}\includegraphics[width=0.45\textwidth]{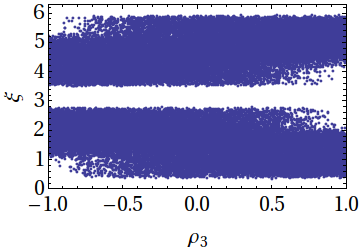}}
    \\
\vspace{-5pt}
  \caption{Correlations between parameters in the Higgs sector. Results of scanning for $w = 300$ GeV, 
  with ranges of parameters defined 
  by eqs. (\ref{ll}-\ref{newlimits}). \label{scan1c}}
\end{figure}

$\bullet$ There is a correlation between a sign of $\rho_2$ (but not of $\rho_3$) and the value of $\xi$ 
as presented in Fig. \ref{r2xi} and Fig. \ref{r3xi}, respectively.
This correlation is related to the positivity of $M_{h_2}^2$ -- by taking a wrong assignment of $(\rho_2, \xi)$ 
pair, e.g. $\pi/2 < \xi < 3\pi/2$ and $\rho_2>0$, we end up with negative $M_{h_2}$.\\

$\bullet$ $\xi$ was initially varied in range $[0,2 \pi]$. We found that
there is a  symmetry in the planes for reflection with respect to  $\xi \sim \pi$, as seen in 
Figs. \ref{L1xi},  \ref{r2xi} and Fig. \ref{r3xi}. 
Therefore, remaining analysis in this paper is limited to values of $\xi\in[0,\pi]$ without affecting the results. 

\begin{figure}[H]
\vspace{-10pt}
  \centering
  \subfloat[$(\lambda_{s1}, M_{h_2})$]{\label{h2ls1}\includegraphics[width=0.45\textwidth]{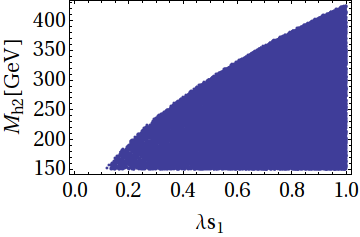}}\quad
    \subfloat[$(\lambda_{s1}, M_{h_3})$]{\label{h3ls1}\includegraphics[width=0.45\textwidth]{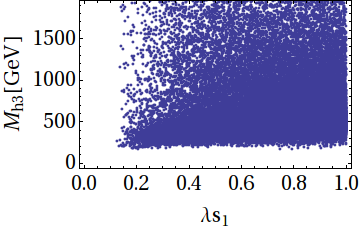}} \\
\vspace{-5pt}
  \caption{Correlations between parameters in the Higgs sector. Results of scanning for $w = 300$ GeV, 
  with ranges of parameters defined by  eqs. (\ref{ll}-\ref{newlimits}). \label{scan2}  }
\end{figure}

\begin{figure}[H]
\vspace{-10pt}
  \centering
  \subfloat[$(\rho_2, M_{h_2})$]{\label{h2r2}\includegraphics[width=0.45\textwidth]{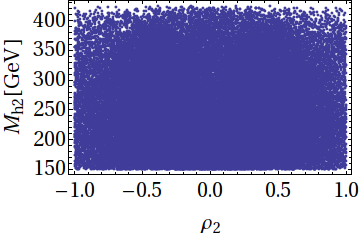}}\quad
    \subfloat[$(\rho_2, M_{h_3})$]{\label{h3r2}\includegraphics[width=0.45\textwidth]{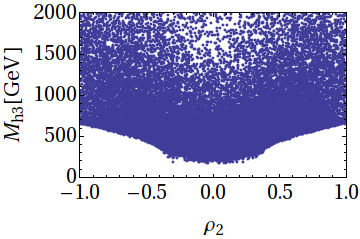}} \\
\vspace{-5pt}
  \caption{Correlations between parameters in the Higgs sector. Results of scanning for $w = 300$ GeV, 
  with ranges of parameters defined by  eqs. (\ref{ll}-\ref{newlimits}).   \label{scan3}  }
\end{figure}

\begin{figure}[H]
\vspace{-10pt}
  \centering
    \subfloat[$(\xi, M_{h_2})$]{\label{h2xi}\includegraphics[width=0.45\textwidth]{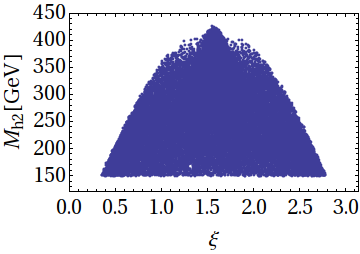}}\quad
    \subfloat[$(\xi, M_{h_3})$]{\label{h3xi}\includegraphics[width=0.45\textwidth]{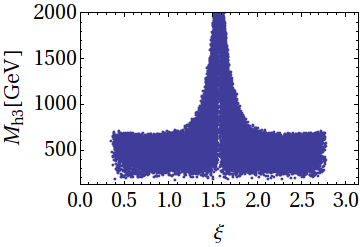}} \\
\vspace{-5pt}
  \caption{Correlations between parameters in the Higgs sector. Results of scanning for $w = 300$ GeV, with ranges of parameters 
  defined by eqs. (\ref{ll}-\ref{newlimits}). \label{scan4}}
\end{figure}

In figures  \ref{scan2}-- \ref{scan4} masses of Higgs particles as a function of some parameters are shown.
In our model, after we expressed the terms $\kappa_i$ in terms of $\rho_i$, we have two mass scales ($v$ and $w$),
and thus the masses of the Higgs particles $h_1,h_2,h_3$ would be given by such values modulo mixing effects.
This can be seen by taking the trace of the mass matrix (eq. \ref{massneut}), which is given by the sum of eqs. (\ref{eq25}), (\ref{eq28})
and (\ref{matrixel}) and it
is also equal to the sum of the mass squared. 
The values of the masses $h_2,h_3$ will get closer or depart from the mass scales $v,w$ depending on the size of the
mixing entries of the mass matrix.

$\bullet$ Fig. \ref{scan2} displays $M_{h_2,h_3}$ versus $\lambda_{s1}$. We can notice that the 
dependence of $M_{h_{2}}$ on parameter $\lambda_{s1}$ reflects the dependence 
on $\Lambda_1$,  which governs  the mixing in the neutral sector (elements 12 and 13 of 
mass matrix $M_{mix}^2$). From Fig. \ref{ls1L1} it is clear  that larger $|\Lambda_1|$ is 
possible for larger $\lambda_{s1}$. Then, the maximum allowed  value of $M_{h_2}$ is related 
to the  perturbativity condition imposed over $\lambda_{s1}$ : 
for $\lambda_{s1} = 0.2$ we can expect masses in range 150 $< M_{h_{2}} < $200 GeV, while for $\lambda_{s1} = 1$ 
the upper limit goes up to about 430 GeV. On the other hand, the allowed values for the mass of $h_3$ are 
higher than for $h_2$ , 170 GeV $< M_{h_{3}} < $O(10 TeV), and are almost
independent of $\lambda_{s1}$, see Fig. 4b for the mass of $h_3$ up to mass 2000 GeV..

$\bullet$  Fig. \ref{scan3} displays $M_{h_2}$, $M_{h_3}$ versus $\rho_{2}$. 
Now the allowed range for the mass of $h_2$ is almost independent of $\rho_2$ and is given by $ 150 < M_{h_2} < 430$ GeV, 
 while the allowed masses for $h_3$ go from
 $ 170 < M_{h_3} < 2000$ GeV for $\rho_2=0$, and are reduced to
  $ 600 < M_{h_3} < 2000$ GeV for $\rho_2= \pm 1$.
Notice the seagull-like shape for the lower 
  limit for $M_{h_3}$, but not for $M_{h_2}$.

 $\bullet$ Fig. \ref{scan4} displays $M_{h_2}$, $M_{h_3}$ versus $\xi$. Here we observe a symmetry for 
 reflection at $\xi \sim\pi/2$. The allowed range, which
 is $ 150 < M_{h_2} < 200$ GeV for $\xi=0.5$, extends up to 
  $ 150 < M_{h_2} < 430$ GeV for $\xi= 1.6$.
Very high mass values for $h_3$ can be obtained for $\xi\sim \pi/2$ (up to $2$ TeV). The trace of the mass 
matrix also help us to understand the larger value of $M^2_{h3}$ for values of $\xi\to \pi/2$,
which comes essentially from the factor $\frac{1}{\cos\xi}$ that appears in eq.(\ref{eq28}). \\
 
\subsection{Parameter space in the inert sector \label{ssec-cordm}}

As discussed in section \ref{mass-neutral-intro}, the masses  of $Z_2$-odd particles are given by a  separate
set of parameters than those of $Z_2$-even particles, which were analyzed in the previous subsection. Here for the inert sector,  three quartic parameters, $\lambda_{3,4,5}$, and one quadratic parameter $m_{22}^2$, are relevant. The remaining 
quartic parameter, $\lambda_2$, appears only in the quartic interaction of $Z_2$-odd particles and is 
therefore not constrained by the analysis of the mass spectrum. However, we expect that -- as in the IDM -- 
combined unitarity, perturbativity and global minimum conditions may provide constraints for this, 
otherwise  practically unlimited, parameter \cite{Swiezewska:2012ej}.

The masses of $Z_2$-odd scalars, and therefore parameters of the potential given by relations (\ref{lam5IDM}) 
and (\ref{relIDM}), are already constrained by experimental and theoretical results.

\begin{enumerate}
\item The LEP studies of invisible decays of $Z$ and $W^\pm$ gauge bosons 
require that there is no decay of $W^\pm$ or $Z$ into inert particles, which gives the following limits \cite{Cao:2007rm,Lundstrom:2008ai}:
\be 
M_{H^\pm} + M_{H,A} > M_{W^\pm}, \quad M_{H} + M_{A} > M_Z , \quad 2M_{H^\pm} > M_Z.
\ee

\item Searches for charginos and neutralinos at LEP have been translated into limits of region of 
masses in the IDM \cite{Lundstrom:2008ai} excluding 
\be 
M_A - M_H > 8 \g  \mbox{ if } M_H < 80 \g \wedge M_A < 100 \g.
\ee
We shall adopt the same limit for inert particles in the studied cIDMS.

\item Note that, as in the IDM, the value of $M_{H^\pm}$ provides limits for $m_{22}^2$, 
which is not constrained by the extremum conditions. Demanding that $M_{H^\pm}^2 > 0$ results 
in $m_{22}^2 < \lambda_{3} v^2$, which for discussed range of $-1\leq\lambda_{3}\leq1$ reduces 
to $m_{22}^2 < v^2$. This constraint is modified by taking account of the "model-independent" 
limit from LEP for the charged scalar mass \cite{Heister:2002ev}:
\be 
M_{H^\pm} > 70-90 \g \Rightarrow m_{22}^2 \lesssim 5 \cdot 10^4 \g^2
\ee
 
Fig. \ref{m222Mch} shows the correlation between the charged-scalar mass and $m_{22}^2$. 
Large values of $M_{H^\pm}$ correspond to large values of $-m_{22}^2$.
\begin{figure}[H]
\centering
\vspace{-10pt}
\includegraphics[width=0.45\textwidth]{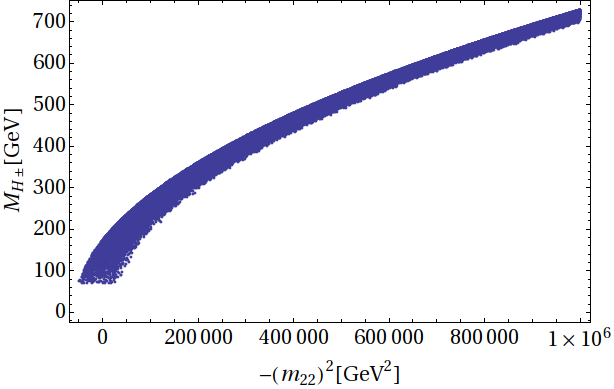}
\vspace{-5pt}
  \caption{Charged scalar mass $M_{H^{\pm}}$ as a function  of $m_{22}^2$.  \label{m222Mch}}
\end{figure}

\item Mass splittings between the $Z_2$-odd particles are given by combinations of 
$\lambda_4$ and $\lambda_5$, 
which are constrained by the perturbativity conditions. If we demand that $|\lambda_{3,4,5}|<1$ 
then in the heavy mass regime all particles
will have similar masses, as they are all driven to high scales by the value of $-m_{22}^2$ (\ref{relIDM}). 
This is visible in Fig. \ref{MHMAMch}. Notice that mass splitting of the order of 200 GeV is  allowed  only for the lighter particles.

\begin{figure}[H]
\vspace{-10pt}
  \centering
    \subfloat[$(M_A, M_H)$]{\includegraphics[width=0.45\textwidth]{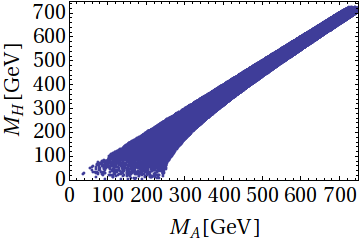}}\quad
    \subfloat[$(M_{H^{\pm}}, M_H)$]{\includegraphics[width=0.45\textwidth]{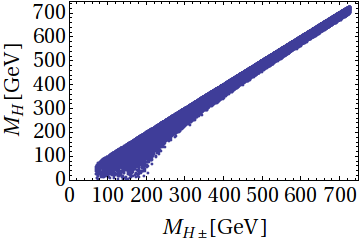}} \\
\vspace{-5pt}
  \caption{(a) Relation between $M_H$ and $M_A$. (b) Relation between $M_{H}$ and $M_{H^{\pm}}$. Both correlations 
  for random scanning with $|\lambda_{3,4,5}|<1$ and $|m_{22}^2| < 10^{6}\g^2$.\label{MHMAMch}}
\end{figure}
\item Electroweak precision measurements provide strong constraints for New Physics beyond the SM.   
In particular, additional particles may introduce important radiative corrections to gauge boson propagators. 
These corrections can be parameterized by the oblique parameters $S$, $T$ and $U$. The value of these parameters 
will be influenced both by the presence of extra (heavy) Higgses present in the cIDMS and by inert particles 
$H^\pm$, $H$ and $A$. $T$ is sensitive to the isospin violation, i.e. it measures the difference between the new physics contributions of neutral and charged current processes at low energies, while $S$ gives new physics contributions to neutral current processes at different energy scales. $U$ is generally small in New Physics models. The latest values of the oblique parameters, determined from a fit with reference mass-values of top and Higgs boson $M_{t,ref}=173 \g$ and $M_{h,ref}=125 \g$ are \cite{Baak:2014ora}:
\begin{equation}
S = 0.05\pm0.11, \quad T = 0.09\pm0.13, \quad U = 0.01\pm0.11.
\end{equation}
In our work we have checked the compatibility of our benchmark points with the 3$\sigma$ bounds on $S$ and $T$, following the method described in \cite{Grimus:2008nb}. For detailed formulas see Appendix \ref{app-stu}. Specific values for given sets of parameters are presented in Table \ref{tab-stu} in Appendix \ref{app-rggdm}. In general, we took the IDM results as the guidance points for our analysis, and found that indeed the cIDMS represents the same behaviour: additional heavy particles, including the heavy Higgses, can be accommodated in the model without violating EWPT constraints.
\item Measurements of invisible decays of the SM-like Higgs at the LHC set very strong constraints on 
Higgs-portal type of DM models [see e.g. \cite{Djouadi:2012zc} and detailed use of constraints in 
\cite{Krawczyk:2013jta} for the IDM, or \cite{Keus:2014jha} for the 3HDM]. 
In general, a DM candidate with mass below approximately 53 GeV annihilating mainly into $b\bar{b}$ 
through the Higgs exchange cannot be in agreement with 
the LHC limits and relic density constraints.  The remaining region, $53 \g \lesssim M_H  \lesssim 62.5 \g$,
corresponds to the Higgs-resonance, 
and the tree-level behaviour is roughly the same in all Higgs-portal-type DM models.
In principle, calculations in this region require loop corrections both for the annihilation cross-section, 
and the scattering cross-section, which is beyond the scope of this work. Therefore, in our analysis we will 
focus on $M_H>M_{h_1}/2$, and comment on the region $M_H<M_{h_1}/2$ in sections \ref{sec-lhc} and \ref{sec-dm} for completeness.

\item For $M_{H}>M_{h_1}/2$, where $h_1$ is the SM-like Higgs particle, all invisible decay channels are closed and 
the most important LHC constraint is now the measured value of $h\to \gamma \gamma$ signal strength, which will be discussed
in detail in the next section.
\end{enumerate}

Further constraints for the DM candidate $H$ come obviously from astrophysical measurements of DM relic density,
and direct and indirect detection. Those will be discussed in section \ref{sec-dm}.

\section{LHC constraints on Higgs parameters in the cIDMS \label{sec-lhc}}

\subsection{Higgs signal strength in the cIDMS}

Further constraints on the parameters of our model (cIDMS) can be obtained by comparing the
light Higgs signal ($h_1$), and the one arising from the SM, with the LHC results. 
This  is done by introducing  the following signal strength:

\begin{equation}
 \mathcal{R}_{XX} = \frac{ \sigma( gg\to h_1 ) }{ \sigma( gg\to \phi_{SM} ) }  
                \frac{\text{BR}(h_1 \to XX) }{\text{BR} (\phi_{SM} \to XX) } ,
\end{equation}
for $X=\gamma, Z,...$,  assuming the gluon fusion is the dominant Higgs production channel at the LHC
and  the narrow-width approximation.
 The expression for
$\mathcal{R}_{XX}$ reduces to:
\begin{equation}\label{Rxx}
 \mathcal{R}_{XX} = \frac{ \Gamma(h_1 \to gg) }{ \Gamma(\phi_{SM} \to gg ) } \, 
                \frac{\text{BR}(h_1 \to XX)}{\text{BR}(\phi_{SM} \to XX)} .
\end{equation}

In our model  the couplings of the lightest Higgs particle ($h_1$) with 
vector bosons and top quark  get modified,   as compared with the SM,
only by a factor $R_{11}$ (where $R_{11}$ is the (11) element of $R^{-1}$ defined by (\ref{r11})).
Thus we can write the Higgs ($h_1$) decay width into gluons  as follows:
 \begin{equation}
       \Gamma(h_1 \to gg)= R_{11}^2 \Gamma(\phi_{SM} \to gg ).
 \end{equation}
Similarly,  for the Higgs boson decay into vector bosons ($V=Z,W$) we have
 \begin{equation}
  \Gamma(h_{1} \to VV^{*})= R_{11}^{2} \Gamma(\phi_{SM} \to VV^{*} ).
 \end{equation}

The one-loop coupling of $h_1$ to photons receives contributions mainly
from the W boson and top quark, as well as the charged scalar $H^\pm$ from the inert sector, so 
the amplitude can be written as\footnote{See Appendix \ref{htoVV0} and references therein for more details.}:
\begin{equation}
 A(h_1\to \gamma \gamma) = R_{11} ( A^{SM}_W+ A^{SM}_t) + A_{H^\pm},
 \end{equation}
 and similar expression for the amplitude {$\cal A$} describing  $ h_1 \to Z \gamma $, see Appendix A.

{ Therefore, the decay widths into two photons and into a photon plus a $Z$ boson,  
are given, respectively, by
\begin{equation}
 \Gamma (h_1\to \gamma \gamma) = R_{11}^2  |1+ \eta_{1}|^2\Gamma (\phi_{SM} \to \gamma \gamma),
\end{equation}
\begin{equation}
 \Gamma(h_1\to Z\gamma)=R_{11}^2 |1+\eta_{2}|^{2}\Gamma(\phi_{SM}\to Z\gamma),
\end{equation}}
where
\begin{eqnarray}
\eta_{1}= \frac{ g_{h_1 H^+ H^-} v } { 2 R_{11} M^2_{H^{\pm}}}\left(\frac{A_{H^\pm}}{A^{SM}_W+ A^{SM}_t}\right) ,\ \
\eta_{2}=\frac{ g_{h_{1}H^{+}H^{-}} v}{2R_{11}M_{H^{\pm}}^{2}}\left(\frac{\mathcal{A}_{H^{\pm}}}{\mathcal{A}_{W}^{SM}+\mathcal{A}_{t}^{SM}}\right).
\end{eqnarray}
The triple coupling $\lambda_{h_1 H^+ H^-}$ is given by
\begin{equation}\label{lhchch}
 g_{h_1 H^+ H^-}=v \lambda_{3} R_{11},
\end{equation}
meaning it is also modified with respect to the IDM by a factor of $R_{11}$.

In the total width of the SM  Higgs boson we can neglect 
 the contributions coming from the Higgs decay into $Z\gamma$ and $\gamma\gamma$.\footnote{Bear in mind that this approximation is established in order to obtain
some analytical expressions for the corresponding ratios, $R_{\gamma\gamma}$, $R_{Z\gamma}$ and $R_{ZZ}$
whose results will guide our dark matter analysis.} 
 The total Higgs decay width in the cIDMS can be significantly modified with respect to the SM if $h_1$ can decay invisibly into inert particles. The 
partial decay width for the invisible channels $h_1 \to \varphi \varphi$, where  $\varphi=A,H$, is:
\be
 \Gamma_{inv}=\Gamma(h_1\to \varphi\varphi)=\frac{g_{h_1\varphi\varphi}^{2}}{32\pi M_{h_{1}}}\left(1-\frac{4M_{\varphi}^{2}}{M_{h_{1}}^{2}}\right)^{1/2},\label{invdec} 
 \ee
 with 
 \be
 g_{h_1AA}=\lambda^{-}_{345} v R_{11}\ \ \text{and} \ \ g_{h_1HH}=\lambda_{345}vR_{11}.\notag
 \ee
 Therefore, in regions of masses where Higgs-invisible decays could take place,  the total width of the Higgs boson in 
 the cIDMS is given by
 \begin{equation}
  \Gamma_{tot}\approx R_{11}^{2} \Gamma_{tot}^{SM}+\Gamma_{inv}.
 \end{equation}

Finally, the signal strengths from Eq.(\ref{Rxx}) can be written as follows,
\begin{eqnarray}\label{rgc}
&\mathcal{R}_{ZZ}= R_{11}^2\zeta^{-1}, \ \ \mathcal{R}_{\gamma\gamma}= R_{11}^2 |1+ \eta_{1}|^2\zeta^{-1},\ \ 
\mathcal{R}_{Z\gamma}=R_{11}^2 |1+\eta_{2}|^{2}\zeta^{-1},
\end{eqnarray}
where $\zeta$ is defined as
\be
\zeta\equiv1+\frac{\Gamma_{inv}}{R_{11}^2\Gamma_{tot}^{SM}}.
\ee

For the cIDMS case $R_{11} = c_1 c_2$, where $c_{1}=\cos\alpha_1$  and $c_{2}=\cos\alpha_2$ are defined by the rotation angles in the 
scalar sector, Eq.(\ref{rotfull}), and thus
\begin{eqnarray}\label{rgc2}
\mathcal{R}_{ZZ}= c_{1}^{2}c_{2}^2\zeta^{-1}, \quad \mathcal{R}_{\gamma\gamma}= c_{1}^{2}c_{2}^2 |1+ \eta_{1}|^2\zeta^{-1}, \quad  
\mathcal{R}_{Z\gamma}=c_{1}^{2}c_{2}^2 |1+\eta_{2}|^{2}\zeta^{-1}.
\end{eqnarray}
Notice that there is a limit on $\mathcal{R}_{ZZ}$, i.e. $\mathcal{R}_{ZZ}\leq1$. It is not possible to enhance 
this decay with respect to the SM. $\mathcal{R}_{\gamma \gamma}$ and $\mathcal{R}_{Z\gamma}$ can be bigger than 1 if 
there is a constructive interference between the SM and the cIDMS contributions. 

\subsection{Numerical analysis of the Higgs signal strenghts}

 Following the discussion in sections  \ref{ssec-cor} and \ref{ssec-cordm} we scan over parameter space in ranges:
\begin{align}
 &0.2\leq\lambda_{1}\leq0.3,\ \ -1\leq\Lambda_{1},\lambda_{3,4},\rho_{2,3}\leq1, \ \ 0\leq\xi\leq \pi, \ \ \notag\\
 & 0 < \lambda_{s1} < 1,\ \  0<\lambda_2 <1, \ \ -1 < \lambda_{5} < 0, \label{A1param}\\
&-10^{6}(\g)^2<m_{22}^{2} < 5\cdot 10^{4}(\g)^2.  \notag
\end{align}
 with $v=246\,\text{GeV}$ and $w = 300 \g$.

\begin{figure}[H]
\vspace{-10pt}
  \centering
  \subfloat[($R_{\gamma\gamma}$,$R_{Z\gamma}$)]{\label{Fa1-1}\includegraphics[width=0.45\textwidth]{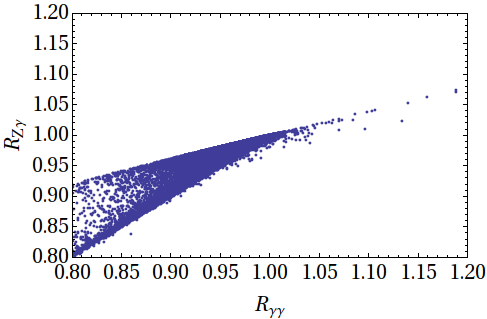}} \quad
   \subfloat[($R_{\gamma\gamma}$,$R_{ZZ}$)]{\label{Fa1-2}\includegraphics[width=0.45\textwidth]{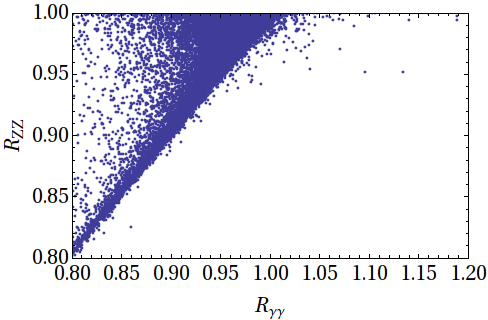}} \\ 
\vspace{-5pt}
  \caption{(a) Correlation between $\mathcal{R}_{\gamma\gamma}$ and $\mathcal{R}_{Z\gamma}$. 
  (b) Correlation between $\mathcal{R}_{\gamma\gamma}$ and $\mathcal{R}_{ZZ}$.}
  \label{Fa1}
\end{figure}

From Fig. \ref{Fa1}  it is clear the ratios $\mathcal{R}_{\gamma\gamma}$, $\mathcal{R}_{Z\gamma}$ 
and $\mathcal{R}_{ZZ}$ can present deviations from the SM value up to $20 \%$. Fig. \ref{Fa1-1} shows the 
correlation between $\mathcal{R}_{\gamma\gamma}$ and
$\mathcal{R}_{Z\gamma}$, while Fig. \ref{Fa1-2} correspond to $\mathcal{R}_{\gamma\gamma}$ and $R_{ZZ}$.  
 If $\mathcal{R}_{\gamma\gamma}<1$ then both $\mathcal{R}_{Z\gamma}$ and $\mathcal{R}_{ZZ}$ are correlated 
with $\mathcal{R}_{\gamma\gamma}$, $\mathcal{R}_{\gamma\gamma}\sim\mathcal{R}_{Z\gamma}$ and $\mathcal{R}_{\gamma\gamma}\sim\mathcal{R}_{ZZ}$.
Notice that there is a possibility of enhancement of both $\mathcal{R}_{\gamma\gamma}$ and $\mathcal{R}_{Z\gamma}$. This is in agreement 
with the IDM, where a correlation between enhancement in $\gamma\gamma$ and $Z\gamma$ channels exists \cite{Swiezewska:2012eh}. 
 Note that the upper limit for $M_{H^{\pm}}$ comes
           from the lower limit for $m_{22}^2$ from set (\ref{A1param})

\begin{figure}[H]
\vspace{-10pt}
  \centering
  \subfloat[($M_{H^{\pm}}$,$R_{\gamma\gamma}$)]{\label{Fa1-3}\includegraphics[width=0.45\textwidth]{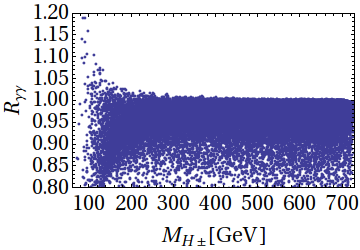}} \quad
   \subfloat[($M_{H^{\pm}}$,$R_{Z\gamma}$)]{\label{Fa1-4}\includegraphics[width=0.45\textwidth]{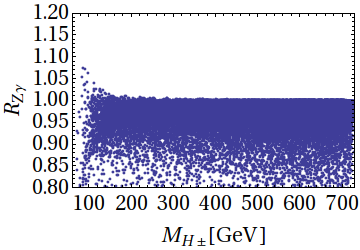}} \\ 
\vspace{-5pt}
  \caption{(a) $\mathcal{R}_{\gamma\gamma}$ as function of $M_{H^{\pm}}$.
           (b) $\mathcal{R}_{Z\gamma}$ as function of $M_{H^{\pm}}$..}
\end{figure}

$\mathcal{R}_{\gamma\gamma}$ and $\mathcal{R}_{Z\gamma}$ as functions of $M_{H^{\pm}}$ are shown in Fig. \ref{Fa1-3} and Fig. \ref{Fa1-4}, respectively
For smaller masses of the charged scalar there is a possibility of enhancement of both $\mathcal{R}_{\gamma\gamma}$ and 
$\mathcal{R}_{Z\gamma}$. For heavier $M_{H^\pm}$ the maximum values tend to the SM value, however deviation up to 20 \%, i.e. 
$\mathcal{R}_{\gamma\gamma,Z\gamma} \approx 0.8$, is possible. Note that the situation is similar to the one from the IDM, 
where significant enhancement, e.g. $\mathcal{R}_{\gamma\gamma} = 1.2$ ,
was possible only if $M_{H^\pm} \lesssim 150 \g$, and for heavier masses $\mathcal{R}_{\gamma\gamma} \to 1$ \cite{Swiezewska:2012eh}.

\begin{figure}[H]
\vspace{-10pt}
  \centering
     \subfloat[($M_{H}$,$R_{\gamma\gamma}$)]{\label{Fa1-5}\includegraphics[width=0.45\textwidth]{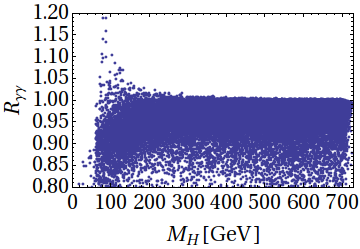}}\quad
    \subfloat[($M_{H}$,$R_{Z\gamma}$)]{\label{Fa1-6}\includegraphics[width=0.45\textwidth]{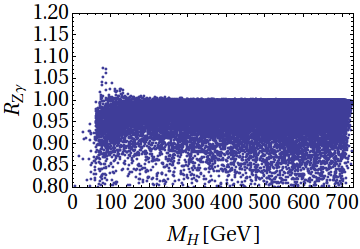}} \\ 
\vspace{-5pt}
  \caption{(a) $\mathcal{R}_{\gamma\gamma}$ as function of $M_{H}$.
           (b) $\mathcal{R}_{Z\gamma}$ as function of $M_{H}$.}\label{RsVsMH}
\end{figure}

\begin{figure}[H]
\vspace{-10pt}
  \centering
\includegraphics[width=0.45\textwidth]{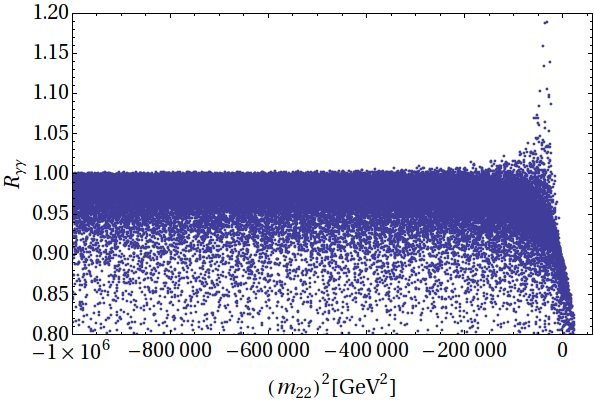}
\vspace{-5pt}
  \caption{$\mathcal{R}_{\gamma\gamma}$ as function of $m_{22}^{2}$.\label{Fa3-1}}
\end{figure}

A similar result is presented in Fig. \ref{Fa3-1}, which depicts $\mathcal{R}_{\gamma\gamma}$ as function of the dimensionful
parameter $m_{22}^{2}$. Significant enhancement is possible only for small values of $|m_{22}^2|$, which correspond to small 
values of $M_{H^\pm}$. For large negative values of $m_{22}^2$, i.e. heavy masses of  all $Z_2$-odd scalars, the preferred value
of $\mathcal{R}_{\gamma\gamma}$ is close to the SM value. Then the heavy particles effectively
decouple from the SM sector and their influence on the SM observables is minimal, as expected. 
This effect it also visible in the IDM.
\subsection{Comment on  invisible Higgs decays}

As mentioned in section \ref{ssec-cordm} measurement of Higgs invisible decays is a powerful tool to
constrain models with additional scalar particles, which couple to the SM-like Higgs $h_1$ and have masses smaller than $M_{h_1}/2$. 
The partial decay width of Higgs into invisible particles, for example a DM candidate from the cIDMS, is given by (\ref{invdec}), and therefore
depends on the DM candidate's mass and its coupling to the Higgs. 

\begin{figure}[H]
\vspace{-10pt}
  \centering
\includegraphics[width=0.4\textwidth]{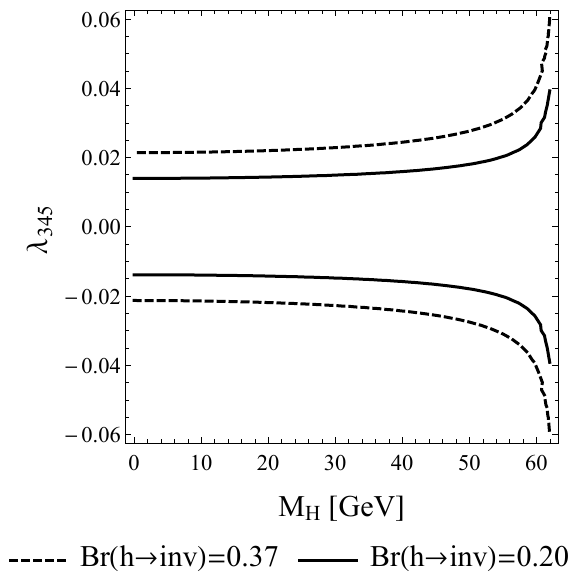}
\vspace{-5pt}
  \caption{Constraints for $\lambda_{345}$ from measurements of Higgs invisible decays branching ratio, with the assumption that only $h_1 \to HH$ channel is open. Solid line: $Br(h\to inv)=0.20$, dashed line: $Br(h\to inv) = 0.37.$  \label{inv5}}
\end{figure}

The cIDMS acts here as a standard Higgs-portal type of DM model and we obtain results known already for the IDM. Figure \ref{inv5} 
shows the permitted range of parameter $\lambda_{345}$, as a function of mass of $M_H$, assuming that $Br(h_1\to inv)$ 
is smaller than $0.37$ (which is the value from ATLAS, denoted by dashed line \cite{atlasbr}) and $0.20$ (which is the value coming from global 
fit analysis, solid line \cite{Belanger:2013xza}).\footnote{This can be treated as a limit for DM-Higgs coupling in the cIDMS, as $g_{H H h_1} = c_1 c_2 \lambda_{345}$, with $c_1 c_2 \approx 0.99$ for all considered SM-like scenarios in the paper.}   
If we demand that $Br(h_1\to inv)<0.37$ allowed region of DM-Higgs coupling is 
$|\lambda_{345} |\lesssim 0.02$ for mass $M_H$ below $\sim$ 30 GeV.
For $Br(h_1\to inv)<0.20$ we obtain $|\lambda_{345}| \lesssim 0.015$. 
This limit will be combined with the relic density 
measurements in section \ref{sec-dm} and it will provide strong constrain, comparable with the one obtained from DM direct 
detection searches, for low DM mass region.

\begin{figure}[H]
\vspace{-10pt}
  \centering
  \subfloat[]{\label{inv1}\includegraphics[width=0.45\textwidth]{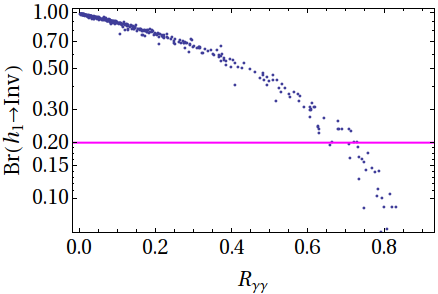}} \quad
  \subfloat[]{\label{inv3}\includegraphics[width=0.45\textwidth]{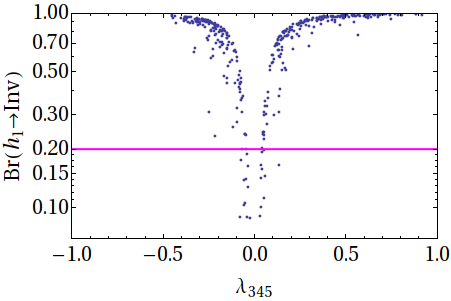}} \\
\vspace{-5pt}
  \caption{(a) $Br(h_1\to inv)$ as a function of $\rgg$. (b) $Br(h_1\to inv)$ 
  as a function of $\lambda_{345}$. In both panels, all the points above $Br(h_1\to inv)=0.2$ 
  are ruled out by current experiment results.}
\end{figure}

In Fig.~\ref{inv1} we see that for a $20\%$ deviation of $R_{\gamma\gamma}$
from (below) the SM model value, the invisible branching ratio is actually
$Br(h_1\to inv)<0.20$. On the other hand, Fig.~\ref{inv3} shows that when the invisible channels are open,
the dimensionless parameter $|\lambda_{345}|$ should be small (as mentioned above) in order to get an invisible branching 
ratio below $20\%$.
In both figures the horizontal line at $Br(h_1\to inv)=0.20$ should be understood as a reference point, so that all the points above 
it are ruled out by current experiment results.

\section{Dark Matter in the cIDMS \label{sec-dm}}

In this section we will discuss properties of DM in the model. Because we can treat the cIDMS as an extension of the IDM, 
we will start with the brief description of DM phenomenology of the later. In both models $H$ is a DM candidate if 
$\lambda_5<0$. In the IDM the DM annihilation channels that are dominant for the DM relic density are 
$HH \to h \to f \bar{f}$ for $M_H \lesssim M_W$ and 
$HH \to WW$ and $HH \to h \to WW$ for $M_H \gtrsim M_W$. If the mass splittings $M_A - M_H$ or $M_{H^\pm} - M_H$ 
are small then the coannihilation channels $H A(H^\pm) \to Z(W^\pm) \to f f'$ also play an important role.

The regions of masses and couplings that correspond to the proper relic density have been studied in many papers 
(see e.g. \cite{Barbieri:2006dq,LopezHonorez:2006gr,Cao:2007rm,Dolle:2009fn,Honorez:2010re,LopezHonorez:2010tb,Sokolowska:2011aa}).
In general, there are four regions of DM mass where the measured relic density can be reproduced:
light DM particles with mass below $10 \textrm{ GeV}$, medium mass regime of $50-80 \textrm{ GeV}$ 
with two distinctive regions: with or without coannihilation of $H$ with the  neutral $Z_2$-odd particle~$A$, 
medium mass region $80-150$ GeV with very large mass splittings, and heavy DM of mass larger than roughly 
$550 \textrm{ GeV}$, where all inert particles have almost degenerate masses and so coannihilation processes 
between all inert particles are crucial. These regions are further constrained or excluded (as it is the case 
with the low DM mass region) by direct and indirect detection experiments, and by the LHC data (see 
e.g. \cite{Krawczyk:2013jta, Goudelis:2013uca,Arhrib:2013ela, Queiroz:2015utg,Garcia-Cely:2015khw} for recent results).

Adding the singlet field $\chi$ changes this picture, although certain properties of the IDM persist. 
In our model there is no direct coupling between the inert doublet $\Phi_2$ and the singlet $\chi$, and the 
only interaction is through mixing of $\chi$ with the first doublet $\Phi_1$. This means, that the 
inert particles' interaction with gauge bosons is like in the IDM, while the inert scalars-Higgs boson
interaction changes with respect to the IDM. The IDM Higgs particle $h$ corresponds 
in our case to $\phi_1$, so $h \to \phi_1$, where $\phi_1 = \beta_1 h_1 + \beta_2 h_2 + \beta_3 h_3$ 
is given by the mixing parameters in (\ref{phi1DM}, with $\beta_1=c_1c_2$ ), and obviously $\sum_{i=1}^3 \beta_i^2 = 1$. 
(The IDM case corresponds to $\beta_{2,3} \to 0$). The important processes for the cIDMS are now:
\begin{eqnarray}
&& HH \to h_i \to f \bar{f}, \quad HH \to h_i \to WW (ZZ), \label{dmdiag}\\
&& HH \to WW,\label{dmgauge}\\
&& H A(H^\pm) \to Z(W^\pm) \to f f', \label{dmcoan}
\end{eqnarray}
with couplings  $g_{h_iHH} = \beta_i g_{h HH}^{\textrm{IDM}}$, \, $g_{h_i f \bar{f}} = \beta_i g_{h f \bar{f}}^{\textrm{IDM}}$, with $g_{h XX}^{\textrm{IDM}}$ being the respective couplings of $h$ to $HH$ and $f \bar{f}$ in the IDM. Following sum rules hold:
\begin{equation}
\sum_{i=1}^3 g_{h_iHH}^2 = (g_{h HH}^{\textrm{IDM}})^2=\lambda_{345}^2,\quad \sum_{i=1}^3 g_{h_i f \bar{f}} = (g_{h f \bar{f}}^{\textrm{IDM}})^2.
\end{equation}

Since both $g_{h_iHH}$ and $g_{h_i f \bar{f}}$ have an extra $\beta_i$ coefficient with respect to the IDM, 
the rate for Higgs-mediated processes (\ref{dmdiag}) will change by $\beta_i^2$. If we are to consider an 
IDM-like case with $\beta_{2,3} \ll \beta_1$ then we could expect to reproduce  results for the IDM. However, 
the interference between diagrams may be in principle important, and as our analysis shows, they 
do influence the results. Notice also, that since CP symmetry is not preserved in this model, 
additional channels like $HH\to h_i \to Z h_j$ can appear and significantly change the relic density value if DM particle is heavy enough.

\subsection{DM constraints}

The masses of inert scalars, including the DM candidate, are constrained in cIDMS, like in the IDM, by 
various experimental limits. Collider constraints for inert particles were discussed in section 
\ref{ssec-cordm}, below we present results and limits from dedicated dark matter experiments.

\begin{enumerate}
\item  We expect the relic density of $H$ to be in agreement with Planck data \cite{Ade:2013zuv}:
\be 
\Omega_{DM} h^2 = 0.1199 \pm 0.0027,
\ee
which leads to the 3$\sigma$ bound:
\be 
 \quad  0.1118<\Omega_{DM}h^2 <0.128. \label{PLANCK_lim}
\ee
If a DM candidate fulfils this requirement, then it constitutes 100 \% of dark matter in the Universe. 
A DM candidate with $\Omega_{DM} h^2 $ smaller than the observed value is allowed, however in this case
one needs to extend the model to have more DM candidates to complement the missing relic density. 
Regions of the parameter space corresponding to value of $\Omega_{DM}h^2$ larger than the Planck upper limit are excluded.
In this work calculation of $\Omega_{DM} h^2$ was performed with an aid of micrOMEGAs 3.5 \cite{Belanger:2013oya}. 
In these calculations all (co)annihilation channels are included, with states with up to two virtual gauge bosons allowed.

\item  
The strongest constraints for light DM annihilating into $bb$ or $\tau\tau$ from indirect detection experiments 
are provided by the measurements of the gamma-ray flux from Dwarf Spheroidal Galaxies by the Fermi-LAT satellite, 
ruling out the canonical cross-section 
$\langle \sigma v\rangle \approx 3\times 10^{-26}~{\rm cm}^3/{\rm s}$ for $M_{DM} \lesssim 100 \mbox{ GeV}$ 
\cite{Ackermann:2015zua}.

For the heavier DM candidates PAMELA and Fermi-LAT experiments provide similar limits of 
$ \langle \sigma v\rangle \approx 10^{-25}~{\rm cm}^3/{\rm s}$ for  $M_{DM}=200 \mbox{ GeV} $ 
in the $bb,\tau\tau$ or $WW$ channels \cite{Cirelli:2013hv}. H.E.S.S. measurements of signal 
coming from the Galactic Centre set limits of $ \langle \sigma v\rangle \approx 10^{-25}-10^{-24}~{\rm cm}^3/{\rm s}$ 
for masses up to TeV scale \cite{Abramowski:2011hc}.

\item Current strongest upper limit on the spin independent (SI) scattering cross section of DM particles on nuclei 
$\sigma_{DM-N}$ is provided by the LUX experiment \cite{Akerib:2013tjd}:
\be  
 \quad \sigma_{DM-N} < 7.6 \times 10^{-46}~{\rm cm}^2 \quad \mbox{for} \quad M_{DM}= 33 \mbox{ GeV}. 
\ee

\end{enumerate}

\subsection{Benchmarks}
In this section we discuss properties of DM for chosen benchmarks in agreement with constraints from LHC/LEP:

\begin{eqnarray}
\textrm{\textbf{A1}: }\; M_{h_1} = 124.83 \g, \; M_{h_2} = 194.46 \g, \; M_{h_3} = 239.99 \g, \\
\textrm{\textbf{A2}: }\; M_{h_1} = 124.85 \g, \; M_{h_2} = 288.16 \g, \; M_{h_3} = 572.25 \g, \\
\textrm{\textbf{A3}: }\; M_{h_1} = 125.01 \g, \; M_{h_2} = 301.41\g, \; M_{h_3} = 1344.01 \g,\\
\textrm{\textbf{A4}: }\; M_{h_1} = 125.36 \g, \; M_{h_2} = 149.89\g, \; M_{h_3} = 473.95 \g.
\end{eqnarray}
By choosing values of $M_{h_1,h_2,h_3}$ we determine parameters from the Higgs sector: $\lambda_1, \lambda_{s1},\Lambda_1,\rho_2,\rho_3,\xi$, as discussed in sec.\ref{ssec-cor}. The corresponding values of parameters of the potential for each benchmark are presented in Appendix \ref{app-bench}.

The above values were chosen to illustrate different possible scenarios:
\begin{itemize}
\item For A1 all Higgs particles are relatively light, although only one, the SM-like Higgs $h_1$, is lighter than $2M_W$.
\item Cases A2 and A3 are similar to A1; the important difference is the value of $M_{h_3}$, which is significantly heavier, and of the order of 500 GeV or 1 TeV, respectively.
\item In scenario A4 there are two Higgs particles that have mass below $2M_W$: $h_1$ (the SM-like Higgs) and $h_2$. 
\end{itemize}
We treat $2M_W$ as the distinguishing value because two Higgs particles of masses smaller than $2 M_{W}$ influence the DM phenomenology by introducing another resonance region in the medium DM  mass regime.

Below we shall discuss properties of DM for the listed benchmark points. In this paper we focus on three different mass regions\footnote{Very light DM particle from the IDM with $M_H \lesssim 10$ GeV is excluded by combined relic density and Higgs-invisible decay limits from the LHC \cite{Krawczyk:2013jta}.}:
\begin{enumerate}
\item light DM mass: $50 \g < M_H < M_{h_1}/2$ with $M_A = M_H + 50 \g, M_{H^\pm} = M_H + 55 \g$,
\item medium DM mass: $M_{h_1}/2 < M_H < M_W$ with $M_A = M_H + 50 \g, M_{H^\pm} = M_H + 55 \g$,
\item heavy DM mass: $M_H \gtrsim 500 \g$ with $M_A = M_{H^\pm} = M_H + 1 \g$,
\end{enumerate}
which are based on studies of the IDM. These mass splittings are in agreement with all collider constraints, including the EWPT limits, for all studied benchmark points (see Table \ref{tab-stu} in Appendix \ref{app-rggdm} for exact values).

We are not going to address the possibility of accidental cancellations in region $M_W <M_H <160-200 \g$ \cite{LopezHonorez:2010tb}, leaving it for the future work. Note however, that this region could in principle be modified with respect to the IDM in benchmarks A2 and A3.
 
\subsection{Light DM}

In this work we define the light DM region as $50 \g < M_H < 62 \g$. As mentioned in 
section \ref{ssec-cordm} and \ref{sec-lhc}, the SM-like Higgs particle can decay invisibly 
into a $HH$ pair (or also into $AA$, if we allow $M_A < M_{h_1}/2$). Measurements of invisible 
decays strongly constrain  the value of the DM-Higgs coupling, which in case of cIDMS is $c_1 c_2 \lambda_{345}$. 

The results presented in this section were obtained for benchmark $A1$. Other benchmarks were also tested 
and they provide no noticeable change in the results. In all considered benchmarks $\beta_1=c_1 c_2 \approx 1$ 
and the main annihilation channel of DM particles is $HH \to h_1 \to b\bar{b}$, regardless of the values of $M_{h_2}$ and $M_{h_3}$. 

\begin{figure}[H]
\vspace{-10pt}
  \centering
\includegraphics[width=0.6\textwidth]{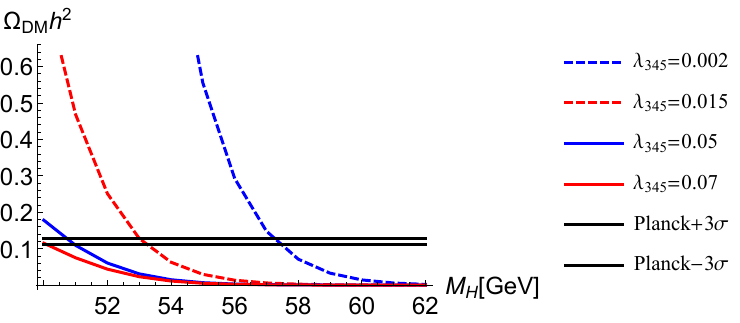}
      
\vspace{-5pt}
  \caption{Values of DM relic density ($\Omega_{DM} h^2$) with respect to DM mass ($M_H$) for chosen values of $\lambda_{345}$ parameter, for benchmark A1. Horizontal lines represent 3$\sigma$  Planck bounds, region above is excluded, in region below additional DM candidate is needed to complement missing DM relic density. Calculations done for $M_{A} = M_{H} + 50 \g, M_{H^\pm}=M_H + 55 \g$, however exact values of those parameters do not influence the output, as the coannihilation effects are surpressed. \label{lightDM}  }
\end{figure}

In the Fig. \ref{lightDM} the relation between $\Omega_{DM}h^2$ and $M_H$ is presented, for a few chosen 
values of $\lambda_{345}$. As discussed before, $|\lambda_{345}|\sim0.015-0.02$ is the boundary value which is in agreement with LHC limits for $Br(h\to inv)$. From Fig. \ref{lightDM} one can see that this value gives the proper relic density for masses of the order of $53 \g$, which is a result that had been previously obtained for One- and Two-Inert Doublet Models \cite{Krawczyk:2013jta,Keus:2014jha}. This value of the coupling for masses below $53 \g$ results in a relic density well above the Planck limits, which leads to overclosing of the Universe. For these smaller masses, to obtain a proper relic density, one needs to enhance the DM annihilation by taking a bigger value of coupling ($|\lambda_{345} \sim 0.05,0.07|$), which at the same time will lead to the enhanced Higgs invisible decays and this is not in agreement with the LHC results. For masses bigger than $53 \g$ coupling corresponding to the proper relic 
abundance gets smaller $(|\lambda_{345}|\sim 0.002)$, fitting into LHC constraints.

 As discussed in section \ref{sec-lhc}, if the Higgs can decay invisibly, its total decay width is strongly affected with respect to 
the SM, and therefore it is not possible to obtain enhancement in the Higgs di-photon decay channel, i.e. $\rgg <1$, see Fig.~\ref{RsVsMH}. 
This was confirmed by a direct check we performed, and the detailed values are presented in the Appendix \ref{app-rggdm} in Table \ref{tab-low}. The maximum allowed value of $\rgg$ for parameters which are in agreement both with the relic density constraints, and with the LHC invisible branching ratio limits, is between $\rgg \approx 0.85-0.91$ for benchmarks A1-A3. It is interesting to note, that for benchmark A4, i.e. the one with two relatively light Higgs particles, the  results are different, here $\rgg$ differs from the SM value by more than 20\%. This is an important difference, because for light DM particles calculation of relic density does not depend on the chosen benchmark.

Similar situation happens with values of $\rzg$, which are close to the SM value for benchmarks A1-A3 (depending on the values of parameters one can obtain both an enhancement or a suppression with respect to $\rzg = 1$), however for benchmark A4 this channel is suppressed by more than 20 \%.

{\DS Recent indirect detection results from Fermi-LAT provide strong constraints for DM candidate annihilating into $b\bar{b}$ pair \cite{Ackermann:2015zua} and are crucial for the low DM mass region. The scalar Higgs-portal type of DM with proper relic density and $M_H \lesssim 53$ GeV is ruled out \cite{Duerr:2015aka}. Heavier masses correspond to the smaller cross-section $\mathcal{O}(10^{-28}-10^{-27}) cm^3/s$. This region is also in agreement with direct detection limits from LUX \cite{Akerib:2013tjd}. Therefore, the only region of low DM mass consistent with all current experimental constraints is the Higgs-resonance region of $53 \g \lesssim M_H \lesssim 62.5 \g$.}

\subsection{Medium DM}

In this section we focus on the medium mass region from the cIDMS, i.e. masses of DM candidate between 
$M_{h_1}/2 \approx 62 \g$ and $M_{W} \approx 83 \g$. 

\begin{figure}[H]
\vspace{-10pt}
  \centering
  \subfloat[A1-A3]{\label{a13}\includegraphics[width=0.45\textwidth]{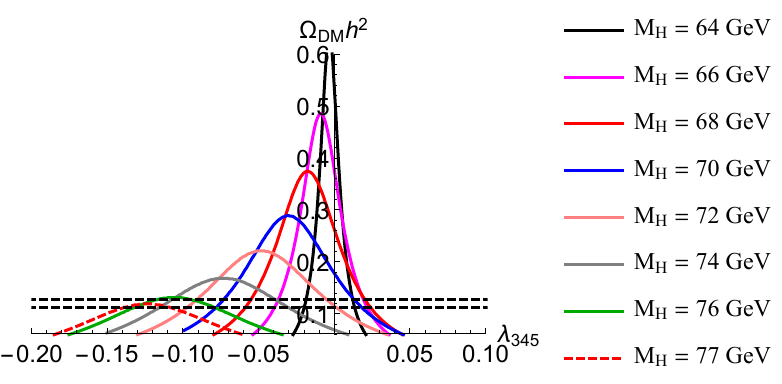}} \quad
    \subfloat[A4]{\label{a4}\includegraphics[width=0.45\textwidth]{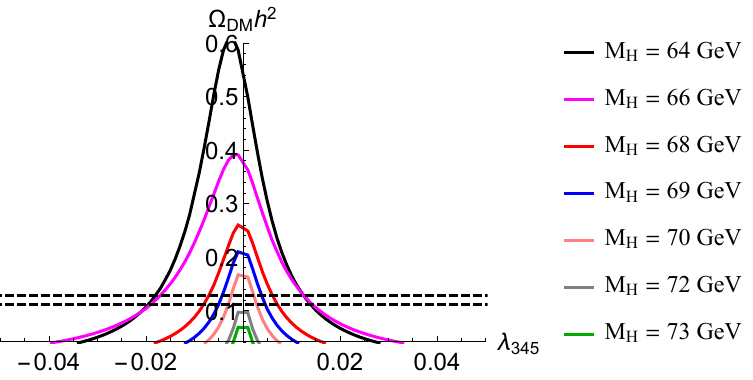}} \\
      \subfloat[IDM]{\label{idm}\includegraphics[width=0.45\textwidth]{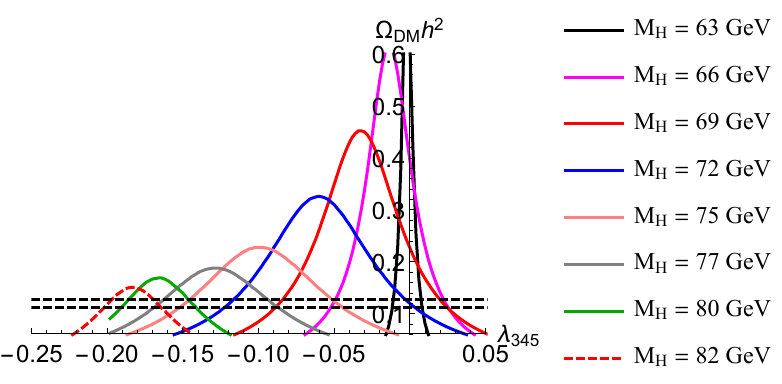}} \quad
      
\vspace{-5pt}
  \caption{Relation between DM relic density $\Omega_{DM}h^2$ and $\lambda_{345}$ for chosen values of $M_H$ for (a) benchmark A2, 
  (b) benchmark A4, (c) the IDM. Horizontal lines represent Planck limits for $\Omega_{DM} h^2 = 0.1199 \pm 3 \sigma$, 
  region above is excluded. Calculations done for $M_{A} = M_{H} + 50 \g, M_{H^\pm}=M_H + 55 \g$, however exact values 
  of those parameters do not influence the output, as the coannihilation effects are surpressed.\label{mid1}  }
\end{figure}

Figures \ref{a13}-\ref{idm} show the behaviour of relic density with respect to $\lambda_{345}$ for masses of 
dark matter candidate changing between $M_{h_1}/2$ and $M_W$, for chosen cIDMS  benchmark points 
A1-A3 (Fig. \ref{a13}) and A4 (Fig. \ref{a4}).  The results for the IDM are well known in the literature; we have included them for comparison in Fig. \ref{idm}.
There is a near-resonance region, $M_H \sim M_h/2$, symmetric around $\lambda_{345} \approx 0$. 
Larger DM masses correspond to greater annihilation into gauge bosons. {\DS The interference between diagrams $HH \to h \to VV \sim \lambda_{345}g$ and $HH \to VV \sim g^2$ depends on the sign of $\lambda_{345}$ and causes }
asymmetry with respect to $\lambda_{345} = 0$. Also, the increased annihilation rate leads to the lowered relic density. 

This behaviour is repeated by benchmark points A1-A3 of cIDMS, where both additional Higgs particles 
are heavier than $2M_W$. However, one can see that the presence of these additional states is non-negligible. 
It is important to stress that even for $\beta_{2,3} \ll \beta_1$, i.e. the case that was supposed to 
be close to the IDM, the impact of three Higgs states on the value of relic density is significant. 
In general, 
the annihilation of DM particles is enhanced and therefore the relic density for a given mass is lower 
with respect to DM candidate from the IDM.
This means, that in the cIDMS for the masses of DM candidate bigger than $79 \g$ relic density is 
below the Planck limit, while for the IDM masses of up to $83 \g$ can be in agreement with the measured value. 

A new phenomena with respect to the IDM can happen if one of the extra Higgs bosons is lighter than $2M_W$, 
which is the case for benchmark A4. As the mass of DM candidate gets closer to this $h_2$-resonance, 
i.e. $M_{DM}\gtrsim 70 \g$, the effective annihilation cross-section increases, resulting in the relic 
density below the observed value. Clearly, the annihilation rate is enhanced and dominated by the Higgs-type 
exchange through $h_2$ (note the symmetric distribution around $\lambda_{345} = 0$), in contrast 
to the previously discussed cases, whereas for the heavier masses the annihilation into gauge bosons 
is starting to dominate, therefore pushing the good region towards negative values of $\lambda_{345}$.

\begin{figure}[H]
\vspace{-10pt}
  \centering
\includegraphics[width=0.8\textwidth]{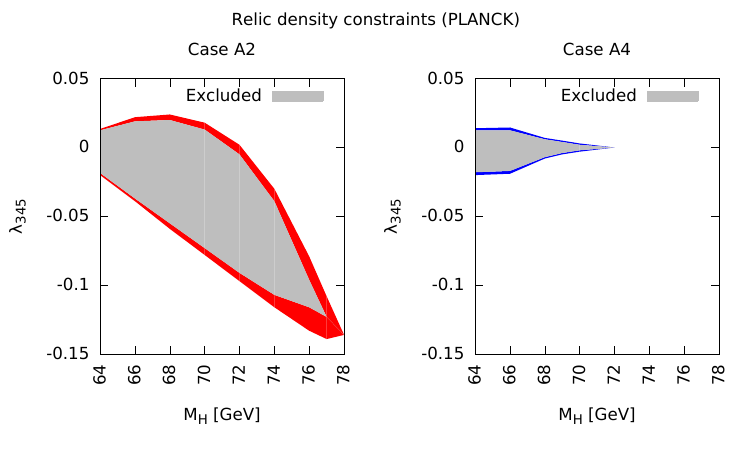}
      
\vspace{-5pt}
  \caption{Relic density constraints on the mass of the DM candidate and its coupling to SM Higgs boson, 
  with the white and gray regions representing too low (not excluded, but an additional DM candidate needs to be added to the model) and too high (excluded) relic abundance, respectively. 
  Red and blue regions corresponds to relic density in agreement with Planck measurements for 
  benchmark A2 and A4, respectively.  \label{mid2}  }
\end{figure}

The difference between benchmarks is even more striking if one studies good regions of relic density in 
the plane $(M_H, \lambda_{345})$, as presented in Fig. \ref{mid2}. For cases A1-A3 the behaviour follows 
that of the IDM, with the corresponding couplings being slightly smaller. Nevertheless, the scenario is 
repeated and one can clearly see the shift towards negative values of $\lambda_{345}$. In case of benchmark 
A4 the situation is completely different; not only the mass range is significantly reduced with respect 
to the previous cases and the IDM, but also the values of coupling are much smaller, concentrated symmetrically around zero.

\begin{figure}[H]
\vspace{-10pt}
  \centering
\includegraphics[width=0.8\textwidth]{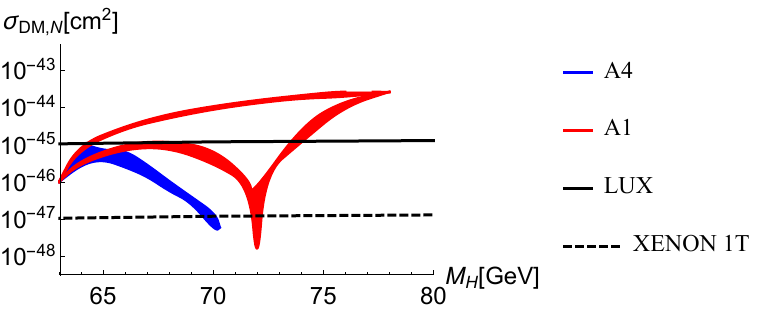}
      
\vspace{-5pt}
  \caption{Direct detection constraints for considered benchmarks (A1-A3: red , A4: blue). All points are in agreement with relic density measurements and collider constraints. Black solid line: upper LUX limit, black dashed line: projected XENON1T limit. \label{ddmid}  }
\end{figure}

The cIDMS, as other scalar DM models, can be strongly constrained by results of direct detection experiments.
The current strongest limits come from LUX experiment, and are presented in Fig. \ref{ddmid}. There are also 
results of calculation of DM-nucleus scattering cross-section, $\sigma_{DM,N}$ for the benchmark points discussed 
in this section. {\DS Red regions denote benchmarks A1-A3, with two separate regions corresponding to two asymmetric branches from Fig.\ref{mid2}. Notice the decrease in cross-section for $M_H \approx 72 \g$, where good relic density is obtained for $\lambda_{345} \approx 0$. Blue region in Fig. \ref{ddmid} corresponds 
to benchmark A4. which is symmetric around $\lambda_{345} =0$ and therefore there is only one branch visible in Fig. \ref{ddmid}}. The difference between those two groups is clear. In case of benchmark A4, the coupling  is usually much smaller than in cases A1-A3, therefore the resulting cross-section will be also 
smaller\footnote{Recall that the DM scattering off nuclei is mediated by the Higgs particles, $h_1,h_2,h_3$, 
therefore the strength of this scattering will directly depend on the value of DM-Higgs couplings.}, 
falling well below the current experimental limits. {\DS However, most of the medium DM mass region is within the reach of future DM direct detection experiments, like XENON1T \cite{Anderson:2011bi} (see Fig.\ref{ddmid}).}

{\DS Recent Fermi-LAT results will constrain the medium mass region, although in the less stringent way than in case of the standard Higgs-portal DM model. Region just above the Higgs-resonance will be excluded by the indirect detection results, as the main annihilation channel for DM candidate is annihilation into $b \bar{b}$ pair of the order of $10^{-26} cm^3/s$. For heavier masses, i.e. $M_H \gtrsim 66 \g$ annihilation into gauge bosons starts to be of the same order as the $b\bar{b}$, and then quickly dominates over all other annihilation channels. The annihilation cross-section gets smaller, of the order of $10^{-27} cm^3/s$. This applies for all studied benchmarks. Therefore, most of the medium DM mass region is in agreement with the current indirect detection limits.}

LHC analysis provides us with further constraints for the studied region. For benchmarks A1-A4 values of $\rgg$ and
$\rzg$ are within the ATLAS \& CMS experimental uncertainties, with the preferred value of $\rgg$ and $\rzg$  below 1. 
The value of these signal strengths depends on the exact values of parameters and an enhancement is possible, 
but not automatic. All values are listed in Table \ref{tab-mid} in Appendix \ref{app-rggdm}. 

Case A4 differs from the other three benchmarks because of the presence of an extra light Higgs particle. 
For points that have good relic density, allowed values of $\rgg$ are close to $\rgg \approx 0.75$, 
with $\rzg$ also below 1, namely $\rzg \sim 0.79$ (see the Table \ref{tab-mid} in Appendix \ref{app-rggdm}).

Recall however, that in contrast with the low DM mass region, here the difference between two groups 
of benchmarks is visible already for calculations of DM relic density.

\subsection{Heavy DM}

In the heavy mass regime all inert particles have similar masses, because of perturbativity limits for 
self-couplings $\lambda_i$. Those masses are driven by the value of $m_{22}^2$, which can reach large negative values. 
Therefore, the mass splittings given by combination of $\lambda_{4,5}$ are small. In this analysis we choose them to be:
\begin{equation}
M_{A} = M_{H^\pm} = M_H + 1 \g.
\end{equation}

\begin{figure}[H]
\vspace{-10pt}
  \centering
\includegraphics[width=0.8\textwidth]{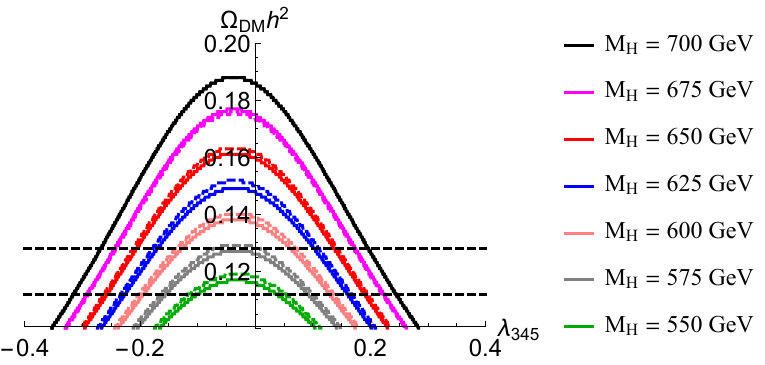}
      
\vspace{-5pt}
  \caption{Heavy DM candidate: relation between relic density and DM-Higgs coupling $\lambda_{345}$ for benchmarks A1 (dashed lines) and A3 (solid lines) for chosen values of $M_H$. Results for A2 and A4 are equivalent to A1. Horizontal lines denote 3$\sigma$ Planck limits.
  \label{heavy}  }
\end{figure}

Fig. \ref{heavy} presents the relation between relic density $\Omega_{DM}h^2$ and DM-Higgs coupling $\lambda_{345}$ for 
benchmarks A1 and A3, for fixed values of DM mass. The difference between A1 and A3 lies in the fact that for 
benchmark A3 there is one very heavy Higgs particle. Note however, that the obtained results are very similar, 
and a very small difference is visible only for masses $M_H \sim 625-650 \g \sim M_{h_3}/2$. For heavy 
masses the 4-vertex annihilation and coannihilation channels into gauge bosons dominate the annihilation 
cross-section, therefore the contribution from additional Higgs states is not nearly as relevant as it was 
for the medium mass region. Therefore we conclude that the presence of heavy Higgs particles of different
masses does not differentiate between the cases.

It is interesting to note, that this region of masses is more similar to the low DM mass region, 
than to the medium mass region. Although all benchmarks result in the very similar values of 
$\Omega_{DM}h^2$, just like for the light DM, there is a difference when it comes to $\rgg$ and $\rzg$. 
Again, for cases A1-A3 the preferred value of $\rgg$ is bigger, this time tending towards the close neighbourhood of 1. For case A4 
resulting values are smaller, close to $\rgg=0.8$. Detailed values are presented in Table \ref{tab-heavy} in Appendix \ref{app-rggdm}.

{\DS Heavy DM candidate from the cIDMS is in agreement with the current DM direct detection limits from LUX, with the average DM-nucleus scattering cross-section of the order of $10^{-46} cm^2$. This region is within range of the future XENON1T experiment. Further constraints for the parameter space of the heavy DM candidate come from the indirect detection experiments, and they provide a complementary way to constrain the region. Analysis performed in \cite{Queiroz:2015utg,Garcia-Cely:2015khw} shows that the H.E.S.S. experiment can already test the parameter space of the IDM, which in the heavy mass region is similar to the cIDMS. Also, the upcoming Cherenkov Telescope Array will be able to probe a significant part of the high mass regime of the models like the IDM or the cIDMS, excluding masses of DM candidate up to 800 GeV.}

\section{Conclusions and Outlook \label{sec-con}}

In this work we have studied the cIDMS -- an extension of the Standard Model, namely a $Z_2$ symmetric Two-Higgs 
Doublet Model with a complex singlet. This model, apart from having a $Z_2$-odd scalar doublet, which may 
provide a good DM candidate, contains a complex singlet with a non-zero complex VEV, which can bring additional 
sources of CP violation. This is a feature that is missing from the IDM.

Within the model different scenarios can be realized. We have focused on the case where the SM-like Higgs particle, 
existence of which has been confirmed by the ATLAS and CMS experiments at the LHC, comes predominantly from the first, 
SM-like doublet, with a small modification coming from the singlet. In addition to the SM-like Higgs there are two other 
Higgs particles, and their presence can strongly influence Higgs and DM phenomenology.

We constrain our model by comparing the properties of the light Higgs particle ($h_1$) from the cIDMS with 
the one arising from the SM. LHC results provide limits for the Higgs-decay signal strengths, in particular $h_1\to\gamma \gamma$. 
There are correlations $\mathcal{R}_{\gamma\gamma}\sim\mathcal{R}_{Z\gamma}$ and $\mathcal{R}_{\gamma\gamma}\sim\mathcal{R}_{ZZ}$. The maximum value for $h_1 \to ZZ$ signal strength is 1.
For smaller masses of the charged scalar there is a possibility of enhancement of both $\mathcal{R}_{\gamma\gamma}$ and 
$\mathcal{R}_{Z\gamma}$. For heavier $M_{H^\pm}$ the maximum values tend to the SM value.
$\mathcal{R}_{\gamma \gamma}$ and $\mathcal{R}_{Z\gamma}$ can be bigger than 1 if 
there is constructive interference between the SM and the cIDMS contributions. 
Notice, that this enhancement is possible simultaneously as in the IDM, 
i.e. there is a correlation between enhancement in $\gamma\gamma$ and $Z\gamma$ 
channels.

The cIDMS can provide a good DM candidate, which is in agreement with the current experimental results. 
The low DM mass region, which we define as masses of $H$ below $M_{h_1}/2$, reproduces behaviour of 
known Higgs-portal DM models, like the IDM. For $M_{H} \lesssim 53$ GeV it is not possible to fulfil LHC constraints for the Higgs 
invisible decay branching ratio and relic density measurements at the same time. 
For $53 \g \lesssim M_H \lesssim 63 \g$ we are in the resonance region of enhanced annihilation with very 
small coupling $\lambda_{345}$ corresponding to proper relic density. This region is in agreement with 
collider and DM direct detection constraints, however we expect the loop corrections to play an important 
role here. It is important to stress that, while DM phenomenology does not depend on the chosen 
benchmark point (A1-A4), there is a difference when it comes to the LHC observables. Values of $\rgg$ 
for benchmark A4 are smaller than in all other cases, being always below 1.

For heavier DM mass, the mere presence of heavier Higgs particles changes the annihilation rate of 
DM particles. Our studies show that the annihilation cross-section is enhanced with respect to the 
IDM and therefore relic density in the cIDMS is usually lower than for the corresponding point in the 
IDM. This is the case both in medium and heavy DM mass region. 

The most striking change with respect to the IDM arises in the relic density analysis with the possibility 
of having an additional resonance region if the mass of one of additional Higgs particles is smaller than $2M_W$. 
For our chosen benchmark points it happens in case A4. Corresponding DM-Higgs couplings, and therefore the 
resulting DM-nucleus scattering-cross-section  constrained by results of direct detection experiments, are 
much smaller for A4 than for other benchmark points. 
This point, however, results in the much smaller values of $\rgg$ and $\rzg$. These values are on the edge 
of 20 \% difference with respect to the SM value, and -- while not being yet excluded by the experiments within 
current experimental errors, they are not favoured. For other studied benchmark points, both relic density 
calculations, and the LHC observables, do not depend very strongly on the exact values of masses of 
Higgs particles. Preferred values of $\rgg$ are of the order of $0.95$.

In the heavy mass region all inert particles are heavier than the particles from the SM sector and 
the impact on the Higgs phenomenology can be minimal. For example, this is the region where $\rgg$ 
is the closest to the SM value. 

Significant modification of our model with respect to the IDM, is the possibility of having 
additional source of CP violation. In a CP-conserving Higgs sector, only real components of Higgs 
multiplets would couple to vector boson pairs (e.g. $h_iZZ$, $h_iW^+W^-$). In the CP-conserving 
2HDM with a real singlet model we would have two CP-conserving neutral states, $h_1,h_2$, 
that couple to $VV$ pair. In a CP-violating Higgs sector, as is the case of cIDMS, 
there is mixing between the real and imaginary parts of Higgs multiplets, resulting in all 
three states $h_1$, $h_2$ and $h_3$ coupling to $VV$ pairs. LHC constraints, which make $h_1VV$ 
couplings so SM-like, suggest the corresponding couplings
of $h_2$ and $h_3$ would be small.

Further CP violating effects may appear in the fermionic sector, when the general Yukawa coupling is 
modified by the CP-violating phases. However,  
by construction only $\Phi_1$ couples to fermions (up-, down-type quarks and charged leptons), 
and such effects are not present, except maybe in the 
neutrino sector.

Therefore we suggest the only possible signal of CP violation would come from  scalar interactions arising 
from the Higgs potential, and in particular those
proportional to parameters $\kappa_2$ or $\kappa_3$. It may be necessary to study the triple interactions 
from the Higgs potential, in order to identify 3-point coupling of the type $h_i h_j h_k$, which would
only appear when there is CP violation present in the model.

The purpose of this paper was to find general properties of the model, which allows for 
additional source of CP violation, at the same time being in agreement with all existing collider data, 
especially on Higgs sector, and dedicated dark matter experiments. Further investigation is needed to 
establish the amount of CP violation provided by the model, which is our plan for the future work.

\subsection*{Acknowledgements}
C.B. was supported by the Spanish grants FPA2014-58183-P, 
Multidark CSD2009-00064 and SEV-2014-0398 (MINECO), 
and PROMETEOII/2014/084 (Generalitat Valenciana).
J.L. Diaz-Cruz acknowledge support from CONACYT-SNI (Mexico). Work of MK, DS
and ND was supported in part by
the grant NCN OPUS 2012/05/B/ST2/03306 (2012-2016). We are thankful 
for valuable discussions with Bogumila Swiezewska. We would like to thank for
very fruitful discussions on the form of the potential to Mario Sampaio, Igor Ivanov 
and Piotr Chankowski. We are also thankful for critical comments from Ilya Ginzburg.

\begin{appendix}

\section{Decays $h\to \gamma\gamma$ and $h\to Z\gamma$}\label{htoVV0}
The decay width, $\Gamma(h\to \gamma\gamma)$, in the IDMS model is given by, \cite{Swiezewska:2012eh,Djouadi:2005gj},
\begin{equation}
 \Gamma(h\to \gamma\gamma)=R_{11}^2 |1+\eta_{1}|^{2}\Gamma(\phi_{SM}\to \gamma \gamma).
\end{equation}
Then the ratio $R_{\gamma\gamma}$ turns out,
\begin{equation}\label{Rgaga}
 \mathcal{R}_{\gamma\gamma}=R_{11}^2 |1+\eta_{1}|^{2},
\end{equation}
where
\begin{equation}\label{eta1}
 \eta_{1}=
 \frac{g_{h_{1}H^{+}H^{-}} v}{2 R_{11} M_{H^{\pm}}^{2}}\left(\frac{A_{H^{\pm}}}{A_{W}^{SM}+A_{t}^{SM}}\right).
\end{equation}
The form factors for this decay are,
\begin{eqnarray}
A_{H^{\pm}}&=&  A_{0}\left(\frac{4M_{H^{\pm}}^{2}}{M_{h_{1}}^{2}}\right),\notag \\
A_{t}^{SM}&=&\frac{4}{3} A_{1/2}\left(\frac{4M^{2}_{t}}{M_{h_{1}}^{2}}\right), \\
A_{W}^{SM}&=& A_{1}\left(\frac{4M^{2}_{W}}{M_{h_{1}}^{2}}\right),\notag
\end{eqnarray}
where,
\begin{eqnarray}
A_{1/2}(\tau)&=&2\tau\left[1+(1-\tau)f(\tau)\right],\notag\\
A_{1}(\tau)&=&-\left[2+3\tau+3\tau(2-\tau)f(\tau)\right],\\
A_{0}(\tau)&=&-\tau\left[1-\tau f(\tau)\right],\notag
\end{eqnarray}
and 
\begin{eqnarray}
f(\tau) =
\left\{
	\begin{array}{ll}
\arcsin^{2}(1/\sqrt{\tau})  & \mbox{for } \tau\geq1 \\
-\frac{1}{4}\left(\log\frac{1+\sqrt{1-\tau}}{1-\sqrt{1-\tau}}-i\pi\right)^2  & \mbox{for } \tau<1.
	\end{array}
\right.
\end{eqnarray}\\
The decay width, $\Gamma(h\to Z\gamma)$, in the IDMS model is given by,
\begin{equation}
 \Gamma(h\to Z\gamma)=R_{11}^2 |1+\eta_{2}|^{2}\Gamma(\phi_{SM}\to Z\gamma)
\end{equation}
and the ratio for this process turns out,
\begin{equation}\label{RZga}
 \mathcal{R}_{Z\gamma}=R_{11}^2 |1+\eta_{2}|^{2},
\end{equation}
where
\begin{equation}\label{eta2}
 \eta_{2}=
 \frac{g_{h_{1}H^{+}H^{-}} v}{2R_{11}M_{H^{\pm}}}\left(\frac{\mathcal{A}_{H^{\pm}}}{\mathcal{A}_{W}^{SM}+\mathcal{A}_{t}^{SM}}\right),
\end{equation}
\begin{eqnarray}
\mathcal{A}_{H^{\pm}}&=&-\frac{(1-2\sin^{2}\theta_{W})}{\cos\theta_{W}}
I_{1}\left(\frac{4M^{2}_{H^{\pm}}}{M_{h}^{2}},\frac{4M^{2}_{H^{\pm}}}{M_{Z}^{2}}\right),\notag \\
\mathcal{A}_{t}^{SM}&=&2\frac{(1-\frac{8}{3}\sin^{2}\theta_{W})}{\cos\theta_{W}}
A_{1/2}^{h}\left(\frac{4M^{2}_{t}}{M_{h}^{2}},\frac{4M^{2}_{t}}{M_{Z}^{2}}\right),\notag \\
\mathcal{A}_{W}^{SM}&=&A_{1}^{h}\left(\frac{4M^{2}_{W}}{M_{h}^{2}},\frac{4M^{2}_{W}}{M_{Z}^{2}}\right),
\end{eqnarray}
\begin{eqnarray}
A_{1/2}^{h}(\tau,\lambda)&=&I_{1}(\tau,\lambda)-I_{2}(\tau,\lambda),\notag\\
A_{1}^{h}(\tau,\lambda)&=&\cos\theta_{W}\left\lbrace 4\left(3-\frac{\sin^{2}\theta_{W}}{\cos^{2}\theta_{W}}\right)I_{2}(\tau,\lambda)+\left[\left(1+\frac{2}{\tau}\right)\frac{\sin^{2}\theta_{W}}{\cos^{2}\theta_{W}}-\left(5+\frac{2}{\tau}\right)\right]I_{1}(\tau,\lambda)\right\rbrace,\notag\\
I_{1}(\tau,\lambda)&=&\frac{\tau\lambda}{2(\tau-\lambda)}+\frac{\tau^{2}\lambda^{2}}{2(\tau-\lambda)^{2}}\left[f(\tau)-f(\lambda)\right]+\frac{\tau^{2}\lambda}{(\tau-\lambda)^{2}}\left[g(\tau^{-1})-g(\lambda^{-1})\right],\notag\\
I_{2}(\tau,\lambda)&=&-\frac{\tau\lambda}{2(\tau-\lambda)}\left[f(\tau)-f(\lambda)\right],
\end{eqnarray}
and
\begin{eqnarray}
g(\tau) =
\left\{
	\begin{array}{ll}
\sqrt{\frac{1}{\tau}-1}\arcsin\sqrt{\tau}  & \mbox{for } \tau\leq1 \\
\frac{\sqrt{1-\frac{1}{\tau}}}{2}\left(\log\frac{1+\sqrt{1-1/\tau}}{1-\sqrt{1-1/\tau}}-i\pi\right)  & \mbox{if } \tau>1.
	\end{array}
\right.
\end{eqnarray}

\section{Oblique parameters \label{app-stu}}

To study contributions to oblique parameters in the cIDMS we use the method presented in \cite{Grimus:2008nb}. There are 6 neutral fields (including a Goldstone boson), related to the physical fields $h_{1-3},H,A$ through:
\begin{equation}
\left(\begin{array}{c}
\varphi_1  + i G^0   \\
H  + i A     \\
\varphi_2  + i \varphi_3   \\
\end{array}\right)= V\left(\begin{array}{c}
G^0 \\
h_1 \\
H   \\
A   \\
h_2 \\
h_3 \\
\end{array} \right),
\end{equation}
The $3\times6$ rotation matrix $V$ is given by
\bea
V=\left(\begin{array}{cccccc}
i & R_{11}         & 0          & 0          & R_{21}          & R_{31} \\
0 & 0              & 1          & i          & 0               & 0 \\
0 & R_{12}+i R_{13}& 0          & 0          & R_{22}+i R_{23} & R_{32}+i R_{33}  \\
\end{array}\right),
\eea
where $R_{ij}$ are the elements of the inverse rotation matrix defined in section \ref{ssec-phys}.

Charged sector contains only a pair of charged scalars $H^\pm$ from doublet $\Phi_2$.

$S$ and $T$ parameters in the cIDMS are given by:

\begin{eqnarray}
 &&T=\frac{g^2}{64 \pi^2 M_W^2 \alpha_{em}}\bigg\{F(M^2_{H^{\pm}}, M_{H}^2)+F(M^2_{H^{\pm}},M_{A}^2)-F(M_H^2,M_{A}^2)\nonumber\\&&-(R_{12}R_{23}-R_{13} R_{22})^2 F(M_{h_1}^2,M_{h_2}^2)\nonumber
\\&&-(R_{12} R_{33} -R_{13} R_{32} )^2F(M_{h_1}^2,M_{h_3}^2)-(R_{22} R_{33} -R_{32} R_{32} )^2F(M_{h_2}^2,M_{h_3}^2)
\\&&+3(R_{11} )^2(F(M_Z^2,M_{h_1}^2)-F(M_W^2,M_{h_1}^2))-3(F(M_Z^2,M_{h_{ref}}^2)-F(M_W^2,M_{h_{ref}}^2))\nonumber
\\&&+3(R_{21} )^2(F(M_Z^2, M_{h_2}^2)-F(M_{W}^2,M_{h_2}^2))+3(R_{31} )^2(F(M_Z^2, M_{h_3}^2)-F(M_{W}^2,M_{h_3}^2))\nonumber
\bigg\}
 \label{T}
\end{eqnarray}
and
\begin{eqnarray}
 &&S=\frac{g^2 }{384 \pi^2 C_w^2}\bigg\{ (2s_w^2-1)^2 G(M_{H^\pm}^2,M_{H^\pm}^2,M_Z^2)+G(M_H^2,M_A^2,M_Z^2)\notag\\
 &&+(R_{12}R_{23}-R_{13}R_{22})^2 G(M_{h_1}^2,M_{h_2}^2,M_Z^2)
+(R_{12} R_{13}-R_{13} R_{32} )^2 G(M_{h_1}^2,M_{h_3}^2,M_Z^2)\notag
\\&&+(R_{22} R_{33}-R_{32} R_{32} )^2 G(M_{h_2}^2,M_{h_3}^2,M_Z^2)
+(R_{11} )^2\widehat{G}(M_{h_1}^2,M_Z^2)
\\&&-\widehat{G}(M_{h_{ref}}^2,M_Z^2)+(R_{21} )^2\widehat{G}(M_{h_2}^2,M_Z^2)+(R_{31} )^2\widehat{G}(M_{h_3}^2,M_Z^2)\notag
\\&&
-2log(M_{H^\pm}^2)+log(M_A^2)+log(M_H^2)+log(M_{h_1})^2-log(M_{h_{ref}})^2+log(M_{h_2})^2+log(M_{h_3})^2 \bigg\},\notag
 \label{S}
\end{eqnarray}

where used functions are defined as:
\bea
F(M_1^2,M_2^2)=\frac{1}{2}(M_1^2+M_2^2)-\frac{M_1^2M_2^2}{M_1^2-M_2^2}log(\frac{M_1^2}{M_2^2}),
\eea
\begin{align}
 G(m_1,m_2,m_3)=&\frac{-16}{3}+\frac{5(m_1+m_2)}{m_3}-\frac{2(m_1-m_2)^2}{m_3^2}\nonumber
\\& +\frac{3}{m_3}  \left[\frac{m_1^2+m_2^2}{m_1-m_2}-\frac{m_1^2-m_2^2}{m_3}+\frac{(m_1-m_2)^3}{3m_3^2}\right]log\frac{m_1}{m_2}+\frac{r f(t, r)}{m_3^3},
\label{G}
\end{align}
The function $f$ of
\begin{equation}
 t\equiv m_1+m_2-m_3 \, \,  and \, \, r\equiv m_3^2-2m_3(m_1+m_2)+ (m_1-m_2)^2 
\end{equation}
is given by
\begin{equation}
f(t, r)=\left\{\begin{array}{cc}
\sqrt{r} \ln|\frac{t-\sqrt{r}}{t+\sqrt{r}}| & r > 0,\\
 0                                         & r = 0,\\
2\sqrt{-r} \arctan\frac{\sqrt{-r}}{t}      & r < 0,
\end{array}
\right. 
\end{equation}
and
\begin{align}
 \widehat{G}(m_1,m_2)=&\frac{-79}{3}+9\frac{m_1}{m_2}-2\frac{m_1^2}{m_2^2}\nonumber
\\& +\left(-10+18\frac{m_1}{m_2}-6\frac{m_1^2}{m_2^2}+\frac{m_1^3}{m_2^3}-9\frac{m_1+m_2}{m_1-m_2}\right)log\frac{m_1}{m_2}\nonumber \\&+(12-4\frac{m_1}{m_2}+\frac{m_1^2}{m_2^2}) \frac{f(m_1,m_1^2-4 m_1 m_2)}{m_2}. \label{'G'}
\end{align}

\section{Benchmarks \label{app-bench}}
Based on analysis done in section \ref{ssec-cor} we propose four benchmark points to be used in DM analysis\footnote{In tables in appendices \ref{app-bench} and \ref{app-rggdm} we are listing parameters with a larger precision to allow the reader to reproduce our results.}. Chosen values of masses of Higgs particles and corresponding parameters are listed in Table \ref{T1}. We also present rotation matrices $R_{Ai}$ for each benchmark. These matrices diagonalize the scalar mass matrix, $M_{mix}^{2}$ in the following way,
\begin{equation}
 \widetilde{M}^{2}=R_{Ai}M_{mix}^{2}R_{Ai}^{T}=\text{diag}(M_{h_{1}}^{2},M_{h_{2}}^{2},M_{h_{3}}^{2}).
\end{equation}

\begin{table}[H]
\centering

\begin{tabular}{|c|c|c|c|c|c|c|c|c|}
\hline
      & $M_{h_{1}}$ & $M_{h_{2}}$ & $M_{h_{3}}$\\ \hline
 $A1$)  & 124.838     &  194.459  & 239.994     \\ \hline
 $A2$)  & 124.852     &  288.161  & 572.235     \\ \hline
 $A3$)  & 125.011     &  301.407  & 1344.01     \\ \hline
 $A4$)  & 125.364     &  149.889  & 473.953     \\ \hline
\end{tabular}\\[5mm]
\centering
\begin{tabular}{|c|c|c|c|c|c|c|c|c|c|c|c|c|c|}
\hline
       &$\lambda_{1}$ &  $\lambda_{s1}$ &  $\Lambda_{1}$  & $\rho_{2}$ &  $\rho_{3}$ & $\xi$  \\ \hline
 $A1$)    & 0.2579     &   0.2241       &  -0.0100        &  0.0881    &  0.1835     & 1.4681         \\ \hline
 $A2$)    & 0.2869     &   0.8894       &  -0.1563        &  0.6892    & 0.6617      & 0.8997          \\ \hline
 $A3$)    & 0.2816     &   0.8423       &  -0.1391        &  0.7010    & -0.5150     & 1.4758      \\ \hline
 $A4$)    & 0.2830     &   0.6990       &   0.0928        &  0.3478    &  0.2900     & 0.4266      \\ \hline
\end{tabular}
\caption{In the first subtable we show the masses of the scalars in GeV. In the
second, the values of Higgs sector dimensionless parameters from the scalar potential are listed.}\label{T1}
\end{table}

\begin{eqnarray}
R_{A1}=\left(
\begin{array}{ccc}
 0.999465 & 0.00682726 & 0.0319988 \\
 -0.0324672 & 0.328031 & 0.944109 \\
 -0.0040509 & -0.944642 & 0.328077
\end{array}
\right).
\end{eqnarray}
\begin{eqnarray}
R_{A2}=\left(
\begin{array}{ccc}
 0.987153 & 0.0555822 & 0.149795 \\
 -0.159095 & 0.255572 & 0.95361 \\
 0.0147203 & -0.965191 & 0.261131
\end{array}
\right).
\end{eqnarray}
\begin{eqnarray}
R_{A3}=\left(
\begin{array}{ccc}
 0.990547 & 0.0252929 & 0.134822 \\
 -0.137173 & 0.186514 & 0.972829 \\
 -0.000540612 & -0.982127 & 0.188221
\end{array}
\right).
\end{eqnarray}
\begin{eqnarray}
 R_{A4}=\left(
\begin{array}{ccc}
 0.90504 & -0.0113276 & -0.425176 \\
 0.424229 & -0.0477451 & 0.904295 \\
 -0.0305436 & -0.998795 & -0.0384057
\end{array}
\right)
\end{eqnarray}

\section{Values of $S, T$ and $\rgg,\rzg$ for studied cases \label{app-rggdm}}

Table \ref{tab-stu} presents values of oblique parameters $S$ and $T$ for chosen values of masses studied in the paper. The 3$\sigma$ bounds are:
\begin{equation}
-0.28 < S <0.38, \quad -0.30 < T < 0.48, \quad -0.32 < U <0.34.
\end{equation}

Table \ref{tab-low}, \ref{tab-mid} and \ref{tab-heavy} contain values of $\rgg$ and $\rzg$ for different values of DM mass, for benchmarks A1-A4. All those points are in agreement with collider and DM constraints.

\begin{table}[h!tb]
\centering
\begin{tabular}{|c|c|c|c|c|c|c|c|c|c|c|c|}
\hline
        & $M_{h_{1}}$ &$M_{h_{2}}$&$M_{h_{3}}$&$ \delta_A$&$\delta_{\pm}$&$M_H$ &$S$&$T$& $3\sigma$ \\ \hline
        &        &        &        &  50  & 55            &  50 & 0.0025 &  0.0050 &   Yes  \\
 $A1$)  &124.838 &194.459 &239.994 &  50  & 55            &  75 & 0.0024 &  0.0051 &   Yes  \\
        &        &        &        &   1  &  1+$\epsilon$ & 600 &-0.0078 &  0.0000 &   Yes  \\ \hline
        &        &        &        &  50  & 55            &  50 & 0.0029 & -0.0378 &   Yes  \\      
 $A2$)  &124.852 &288.161 &572.235 &  50  & 55            &  75 & 0.0028 & -0.0377 &   Yes  \\      
        &        &        &        &  1   &  1+$\epsilon$ & 600 &-0.0075 & -0.0418 &   Yes  \\ \hline
        &        &        &        &  50  & 55            &  50 & 0.0031 & -0.2177 &   Yes   \\      
 $A3$)  &125.011 &301.407 &1344.01 &  50  & 55            &  75 & 0.0030 & -0.2176 &   Yes   \\      
        &        &        &        &  1   &  1+$\epsilon$ & 600 &-0.0072 & -0.2228 &   Yes   \\ \hline
        &        &        &        &  50  & 55            &  50 & 0.0027 & -0.1968 &   Yes   \\      
 $A4$)  &125.364 &149.889 &473.953 &  50  & 55            &  75 & 0.0026 & -0.1967 &   Yes   \\      
        &        &        &        &  1   &  1+$\epsilon$ & 600 &-0.0077 & -0.2019 &   Yes   \\ \hline
\end{tabular}
\caption{Values of oblique parameters $S$ and $T$ for benchmark points $A1-A4$ and chosen masses of inert scalars. All studied cases are in agreement with EWPT constraints.}\label{tab-stu}
\end{table}

\begin{table}
\begin{minipage}{3in}
\centering
\textbf{Benchmark A1}\\[2mm]
\begin{tabular}{|c|c|c|c|}
\hline
$M_H$~(GeV)&$|\lambda_{345}|$&$\rgg$&$\rzg$\\ \hline
50&0.015&0.8770&0.9365\\ \hline

53&0.015&0.8826&0.9405\\ \hline

56&0.015&0.8886&0.9449\\ \hline

59&0.015&0.8952&0.9503\\ \hline

50&0.002&0.9014&0.9596\\ \hline

53&0.002&0.9045&0.9611\\ \hline

56&0.002&0.9073&0.9624\\ \hline

59&0.002&0.9100&0.9636\\ \hline

50&0.001&0.9020&0.9601\\ \hline

53&0.001&0.9050&0.9615\\ \hline

56&0.001&0.9078&0.9627\\ \hline

59&0.001&0.9104&0.9639\\ \hline
\end{tabular}
\end{minipage}
\begin{minipage}{3in}
\centering
\textbf{Benchmark A2}\\[2mm]
\begin{tabular}{|c|c|c|c|}
\hline
$M_H$~(GeV)&$|\lambda_{345}|$&$\rgg$&$\rzg$\\ \hline
50&0.015&0.8556&0.9136\\ \hline

53&0.015&0.8610&0.9175\\ \hline

56&0.015&0.8668&0.9218\\ \hline

59&0.015&0.8733&0.9270\\ \hline

50&0.002&0.8793&0.9361\\ \hline

53&0.002&0.8823&0.9375\\ \hline

56&0.002&0.8851&0.9388\\ \hline

59&0.002&0.8877&0.9400\\ \hline

50&0.001&0.8799&0.9366\\ \hline

53&0.001&0.8829&0.9379\\ \hline

56&0.001&0.8856&0.9392\\ \hline

59&0.001&0.8882&0.9403\\ \hline
\end{tabular}
\end{minipage}\\[2mm]
\begin{minipage}{3in}
\centering
\textbf{Benchmark A3}\\[2mm]
\begin{tabular}{|c|c|c|c|}
\hline
$M_H$~(GeV)&$|\lambda_{345}|$&$\rgg$&$\rzg$\\ \hline
50&0.015&0.8615&0.9199\\ \hline

53&0.015&0.8670&0.9239\\ \hline

56&0.015&0.8728&0.9281\\ \hline

59&0.015&0.8794&0.9334\\ \hline

50&0.002&0.8854&0.9426\\ \hline

53&0.002&0.8884&0.9440\\ \hline

56&0.002&0.8912&0.9453\\ \hline

59&0.002&0.8939&0.9465\\ \hline

50&0.001&0.8860&0.9430\\ \hline

53&0.001&0.8890&0.9444\\ \hline

56&0.001&0.8917&0.9456\\ \hline

59&0.001&0.8943&0.9468\\ \hline
\end{tabular}
\end{minipage}
\begin{minipage}{3in}
\centering
\textbf{Benchmark A4}\\[2mm]
\begin{tabular}{|c|c|c|c|}
\hline
$M_H$~(GeV)&$|\lambda_{345}|$&$\rgg$&$\rzg$\\ \hline
50&0.015&0.7192&0.7679\\ \hline

53&0.015&0.7238&0.7712\\ \hline

56&0.015&0.7287&0.7748\\ \hline

59&0.015&0.7341&0.7792\\ \hline

50&0.002&0.7391&0.7870\\ \hline

53&0.002&0.7417&0.7881\\ \hline

56&0.002&0.7440&0.7892\\ \hline

59&0.002&0.7463&0.7902\\ \hline

50&0.001&0.7396&0.7872\\ \hline

53&0.001&0.7421&0.7884\\ \hline

56&0.001&0.7445&0.7894\\ \hline

59&0.001&0.7466&0.7904\\ \hline
\end{tabular}
\end{minipage}
\caption{Low DM mass region: values of $\rgg$ and $\rzg$ for chosen values of $M_H$ and $\lambda_{345}$ for $M_A = M_H + 50 \g, M_{H^\pm} = M_H + 55 \g$. Points listed above correspond to DM relic density in agreement with Planck results. Values of $\rgg$ and $\rzg$ do not depend on the sign of $\lambda_{345}$. \label{tab-low}}
\end{table}

\begin{table}
\begin{minipage}{3in}
\centering
\textbf{Benchmark A1}\\[2mm]
\begin{tabular}{|c|c|c|c|}
\hline
$M_H$~(GeV)&$\lambda_{345}$&$\rgg$&$\rzg$\\ \hline
64&0.0125&0.9116&0.9646\\ \hline

66&0.019&0.9116&0.9646\\ \hline

68&0.02&0.9130&0.9665\\ \hline

70&0.018&0.9149&0.9660\\ \hline

72&-0.097&0.94010&0.9764\\ \hline

74&-0.039&0.9295&0.9719\\ \hline

76&-0.116&0.9458&0.9783\\ \hline

77&-0.123&0.9474&0.9800\\ \hline

78&-0.136&0.9501&0.9800\\ \hline
\end{tabular}
\end{minipage}
\begin{minipage}{3in}
\centering
\textbf{Benchmark A2}\\[2mm]
\begin{tabular}{|c|c|c|c|}
\hline
$M_H$~(GeV)&$\lambda_{345}$&$\rgg$&$\rzg$\\ \hline
64&0.0125&0.8893&0.9410\\ \hline

66&0.019&0.8893&0.9410\\ \hline

68&0.02&0.8906&0.9416\\ \hline

70&0.018&0.8925&0.9424\\ \hline

72&-0.097&0.9179&0.9525\\ \hline

74&-0.039&0.9067&0.9481\\ \hline

76&-0.116&0.9227&0.9544\\ \hline

77&-0.123&0.9242&0.9550\\ \hline

78&-0.136&0.9268&0.9560\\ \hline
\end{tabular}
\end{minipage}\\[2mm]
\begin{minipage}{3in}
\centering
\textbf{Benchmark A3}\\[2mm]
\begin{tabular}{|c|c|c|c|}
\hline
$M_H$~(GeV)&$\lambda_{345}$&$\rgg$&$\rzg$\\ \hline
64&0.0125&0.8954&0.9475\\ \hline

66&0.019&0.8955&0.9475\\ \hline

68&0.02&0.8968&0.9481\\ \hline

70&0.018&0.8987&0.9490\\ \hline

72&-0.097&0.9243&0.9590\\ \hline

74&-0.039&0.9130&0.9546\\ \hline

76&-0.116&0.9290&0.9610\\ \hline

77&-0.123&0.9306&0.9616\\ \hline

78&-0.136&0.9332&0.9626\\ \hline
\end{tabular}
\end{minipage}
\begin{minipage}{3in}
\centering
\textbf{Benchmark A4}\\[2mm]
\begin{tabular}{|c|c|c|c|}
\hline
$M_H$~(GeV)&$\lambda_{345}$&$\rgg$&$\rzg$\\ \hline
64&-0.02&0.7542&0.7936\\ \hline

66&-0.017&0.7546&0.7938\\ \hline

68&0.006&0.7513&0.7925\\ \hline

69&0.004&0.7523&0.7929\\ \hline

70&-0.003&0.7540&0.7937\\ \hline\end{tabular}
\end{minipage}
\caption{Medium DM mass region: values of $\rgg$ and $\rzg$ for chosen values of $M_H$ and $\lambda_{345}$ for $M_A = M_H + 50 \g, M_{H^\pm} = M_H + 55 \g$. Points listed above correspond to DM relic density in agreement with Planck results. \label{tab-mid}}
\end{table}

\begin{table}
\begin{minipage}{3in}
\centering
\textbf{Benchmark A1}\\[2mm]
\begin{tabular}{|c|c|c|c|}
\hline
$M_H$~(GeV)&$\lambda_{345}$&$\rgg$&$\rzg$\\ \hline
550&0&0.9986&0.9989\\ \hline
575&0.2&0.9967&0.9981\\ \hline
575&-0.2&1.0005&0.9995\\ \hline
600&0.23&0.9966&0.9981\\ \hline
600&-0.23&1.0006&0.9995\\ \hline
625&0.25&0.9966&0.99981\\ \hline
625&-0.25&1.0006&0.9995\\ \hline
650&0.28&0.9966&0.9980\\ \hline
650&-0.28&1.0007&0.9996\\ \hline
675&0.3&0.9966&0.9981\\ \hline
675&-0.3&1.0007&0.9996\\ \hline
700&0.33&0.9965&0.9980\\ \hline
700&-0.33&1.0007&0.9996\\ \hline
\end{tabular}
\end{minipage}
\begin{minipage}{3in}
\centering
\textbf{Benchmark A2}\\[2mm]
\begin{tabular}{|c|c|c|c|}
\hline
$M_H$~(GeV)&$\lambda_{345}$&$\rgg$&$\rzg$\\ \hline
550&0&0.9741&0.9743\\ \hline

575&0.2&0.9723&0.9737\\ \hline

575&-0.2&0.9760&0.9750\\ \hline

600&0.23&0.9722&0.9736\\ \hline

600&-0.23&0.9761&0.9751\\ \hline

625&0.25&0.9722&0.9736\\ \hline

625&-0.25&0.9761&0.9751\\ \hline

650&0.28&0.9722&0.9736\\ \hline

650&-0.28&0.9762&0.9751\\ \hline

675&0.3&0.9722&0.9736\\ \hline

675&-0.3&0.9762&0.9751\\ \hline

700&0.33&0.9721&0.9736\\ \hline

700&-0.33&0.9762&0.9751\\ \hline
\end{tabular}
\end{minipage}\\[2mm]
\begin{minipage}{3in}
\centering
\textbf{Benchmark A3}\\[2mm]
\begin{tabular}{|c|c|c|c|}
\hline
$M_H$~(GeV)&$\lambda_{345}$&$\rgg$&$\rzg$\\ \hline
550&0&0.9808&0.9810\\ \hline

575&0.2&0.978982&0.9804\\ \hline

575&-0.2&0.9827&0.9817\\ \hline

600&0.23&0.9789&0.9803\\ \hline

600&-0.23&0.9828&0.9818\\ \hline

625&0.25&0.9789&0.9803\\ \hline

625&-0.25&0.9828&0.9818\\ \hline

650&0.28&0.9788&0.9803\\ \hline

650&-0.28&0.9829&0.9818\\ \hline

675&0.3&0.9789&0.9803\\ \hline

675&-0.3&0.9829&0.9818\\ \hline

700&0.33&0.9788&0.9803\\ \hline

700&-0.33&0.9830&0.9818\\ \hline
\end{tabular}
\end{minipage}
\begin{minipage}{3in}
\centering
\textbf{Benchmark A4}\\[2mm]
\begin{tabular}{|c|c|c|c|}
\hline
$M_H$~(GeV)&$\lambda_{345}$&$\rgg$&$\rzg$\\ \hline
550&0&0.8188&0.8190\\ \hline

575&0.2&0.8173&0.8184\\ \hline

575&-0.2&0.8203&0.8196\\ \hline

600&0.23&0.8172&0.8184\\ \hline

600&-0.23&0.8204&0.8196\\ \hline

625&0.25&0.8172&0.8184\\ \hline

625&-0.25&0.8205&0.8196\\ \hline

650&0.28&0.8172&0.8184\\ \hline

650&-0.28&0.8205&0.8196\\ \hline

675&0.3&0.8172&0.8184\\ \hline

675&-0.3&0.8205&0.8196\\ \hline

700&0.33&0.81714&0.8196\\ \hline

700&-0.33&0.82057&0.81964\\ \hline
\end{tabular}
\end{minipage}
\caption{Heavy DM mass region: values of $\rgg$ and $\rzg$ for chosen values of $M_H$ and $\lambda_{345}$ for $M_A = M_{H^\pm} = M_H + 1 \g$. Points listed above correspond to DM relic density in agreement with Planck results. \label{tab-heavy}}
\end{table}

\end{appendix}

\baselineskip 16pt
\bibliographystyle{unsrt}

\end{document}